\documentclass[useAMS,usenatbib,review]{mn2e}
\usepackage{dblfloatfix}
\usepackage{mathtext,amssymb}
\usepackage{epsfig}
\usepackage{graphics}
\usepackage{url}

 \usepackage{times}

\renewcommand{\vector}[1]{\ensuremath{\mathbf{#1}}}
\renewcommand{\deg}{\ensuremath{\rm ^\circ}}

\newcommand{\magdot}[1]{\ensuremath{#1\!^{\rm m}}}

\newcommand{\Msun}{\ensuremath{\mathrm{M_\odot}}}

\newcommand{\kms}{\ensuremath{\mathrm{km\,s^{-1}}}}
\newcommand{\Gpc}{\ensuremath{\mathrm{Gpc}}}

\newcommand{\pc}{\ensuremath{\mathrm{pc}}}
\newcommand{\mJy}{\ensuremath{\mathrm{mJy}}}

\newcommand{\sbs}{SBS~J1520+530}
\newcommand{\einc}{QSO~J2237+0305}
\newcommand{\pg}{PG~1115+080}
\newcommand{\rxj}{RXJ~1131-1231}
\newcommand{\sdss}{SDSS~0924+0219}

\newcommand{\AAA}{\ensuremath{\rm \AA}}

\renewcommand{\div}{\ensuremath{..}}

\title[Resolving QSO Discs]{Resolving the Inner Structure of QSO Discs by Fold
  Caustic Crossing Events}
\author[Abolmasov \& Shakura]{P.\ Abolmasov$^{1,2}$\thanks{E-mail:
pavel.abolmasov@gmail.com} and N.\ I.\ Shakura$^{1,2}$ \\ 
$^{1}$Sternberg Astronomical Institute, Moscow State University,
  Universitetsky pr., 13, Moscow, 119992, Russia\\
$^{2}$Max Planck Institut f\"{u}r Astrophysik, Karl-Schwarzschild-Str. 1,
85741 Garching, Germany\\
}
\begin{document}

\date{Accepted ---. Received ---; in
  original form --- }

\pagerange{\pageref{firstpage}--\pageref{lastpage}} \pubyear{2009}

\maketitle

\label{firstpage}

\begin{abstract}
Though the bulk of the 
observed optical flux from the discs of intermediate-redshift 
lensed quasars is formed well
outside the region of strong relativistic boosting and light-bending,
relativistic effects have important influence on microlensing
curves. The reason is in the divergent nature of amplification factors
near fold caustics increasingly sensitive to small spatial size
details. Higher-order disc images produced by strong light bending around the
black hole may affect the amplification curves, making a contribution
of up to several percent near maximum amplification.
In accordance with theoretical predictions, some of the observed
high-amplification events possess fine structure. Here we consider three
putative caustic crossing events, one by \sbs\ and two events
 for individual images of the Einstein's cross (\einc). Using
relativistic disc models allows to improve the fits, but the required
inclinations are high, $i\gtrsim 70^\circ$. Such high inclinations
apparently contradict the 
absence of any strong absorption that is likely to arise if a disc
is observed edge-on through a dust torus. Still, the high
inclinations are required only for the central parts of the disc, that
allows the disc itself to be initially tilted by $60\div 90^\circ$ with respect
to the black hole and aligned
toward the black hole equatorial plane near the last stable orbit radius. For
\sbs, an alternative explanation for the observed amplification curve
is a superposition of two subsequent fold caustic crossings. While relativistic disc
models favour black hole masses $\sim 10^{10}\Msun$ (several times
higher than the virial estimates) or small Eddington ratios, this
model is consistent with the observed distribution of galaxies over
peculiar velocities only if the black hole mass is $\lesssim 3\times
10^8\Msun$.
\end{abstract}


\section{Introduction}

The history of gravitational lensing may be traced down to the
beginning of the XXth century and beyond (see
\citet{schmidt_wambsganss} and references therein for a historical
review). First double quasar images produced by strong lenses were
reported in \citet{walsh79}. The first quadruply-lensed quasar, \einc,
was discovered by \citet{huchra}. This object was also the first were
effects of microlensing by the stars of the lensing galaxy were found
\citep{einc_micro}. 

The principal difference between strong lensing and
microlensing is in the angular distance scale set by the Einstein-Chwolson
radius:

\begin{equation}\label{E:thein}
\theta_{Ein} = \sqrt{ \frac{4GM}{c^2} \frac{D_{LS}}{D_{S}
    D_{L}}} \simeq 2.8 \
\sqrt{\frac{M}{\Msun} \frac{D_{LS}}{D_{L}} \frac{1\Gpc}{D_S}} \rm
\mu as
\end{equation}

Here, $M$ is the mass of the lensing object, $D_{L,S, LS}$ are angular
size distances toward the lens (``L''), the source (``S'') and between
the source and the lens (``LS''), $D_{LS} = D_S - D_L \times (1+z_L)/(1+z_S)$
for a flat Universe (see for example \citet{Hogg99}). 
 Dependence on the lens
mass leads to drastically different angular scales associated with
strong lensing by galaxies and galaxy groups (arcseconds) and
microlensing by individual stars (microarcseconds and less). The individual
images formed in the latter case cannot be resolved by contemporary
instrumentation, but their amplification variations 
may be a valuable tool to resolve the
spatial structure of the source \citep{CR84}. Accretion discs around
supermassive black holes at cosmological distances should have
comparable or somewhat smaller angular sizes at
$\lambda \sim 2000\AAA$. Below we will use spatial sizes projected onto the
picture plane at the distance of the source. They differ from the
angular sizes by the dimensional factor of $D_S$:

\begin{equation}\label{E:rein}
r_{Ein} = \sqrt{ \frac{4GM}{c^2} \frac{D_{LS}D_S}{D_{L}}} 
\simeq 4.3\times 10^{16} \
\sqrt{\frac{M}{\Msun} \frac{D_{LS}}{D_{L}} \frac{D_S}{1\Gpc}} \rm
cm
\end{equation}

The main difference between microlensing by stellar mass objects of
our Galaxy
and microlensing effects accompanying strong lensing of quasars is in the
optical depth. Angular distances between individual
lensing bodies in a galaxy scale with the distance $D$ as
$\propto D^{-1}$, while Einstein-Chwolson radii decrease only as
$\propto D^{-0.5}$. If we define the microlensing optical depth as the
total solid angle of the Einstein circles of the stellar population of
a unit solid angle
of the lensing galaxy (cf. \citet{NO84}), it may be written as:

\begin{equation}\label{E:tau}
\tau = \frac{4\pi G\Sigma}{c^2}\frac{D_{LS}D_L}{ D_S} 
 \simeq 0.06 \frac{\Sigma}{100\Msun\pc^{-2}}
\frac{D_{LS}}{D_S} \frac{D_{L}}{1\Gpc} 
\end{equation}

Here, $\Sigma$ is the total surface density of the clumped matter in
the lensing galaxy. Smoothly distributed component (primarily,
unclumped dark matter) contributes only to strong lensing, unless it
creates a strong shear that destroys point-lens degeneracy \citep{NO84}.
Optical depth becomes considerably large ($\tau \gtrsim 0.1$) for
lensed quasars\footnote{Here we do not distinguish between radio-loud
  quasars and
radio-quiet ``quasi-stellar objects'' (QSO) and refer to both object
types as ``quasars'' or QSO.} at gigaparsec distances.

Difference in optical depth makes microlensing effects in distant
lensing galaxies qualitatively different from the rare single- and
double-lens events in our Galaxy. As a background source
moves with respect to a single point-like lens, it has zero
probability to undergo
infinite amplification \citep{paczynski86a}. In this regime,
amplification becomes sensitive to the size and the structure of the
object only if its angular distance from the lens centre becomes comparably
small. 
On the other hand, strong ($\tau \gtrsim 0.1\div 0.5$, depending on the
underlying shear) microlensing
by a population of point masses creates a network of fold caustics
\citep{paczynski86b} where amplification is divergent and behaves as $\propto
d^{-1/2}$, where $d$ is the angular distance toward the fold \citep{CR84}.
In the case of quasar microlensing, every particular
image traverses some fold caustic at a probability of about unity
on a several years' time scale (for the case of \einc, the
relative frequency of high-amplification events was estimated as $\sim
1\rm yr^{-1}$ by \citet{WKR93}, other objects have smaller
relative proper motions). A comprehensive review on microlensing was
made by \citet{wambsganss}.

At present, there are at least three different approaches to quasar
microlensing that may be used to probe the structure of supermassive
black hole accretion discs. First is to gather statistics on
image amplification and to compare single-epoch anomalous fluxes with the
predictions of accretion disc models, as it was done by
\citet{pooley07,bate08,floyd09,blackburne11,jimenez}. 
In the first paper, a huge,
about a factor of $10\div100$, inconsistency was found between the
sizes of accretion discs estimated by microlensing methods and
predictions of the standard accretion disc theory. Partially, this
inconsistency may be attributed to the crude mass estimates applying
broad-band magnitudes with some assumptions on bolometric correction
and accretion efficiency. On the other hand, \citet{morgan10} use
more accurate mass estimates based on the widths of broad emission lines and
yield much better consistency.
Indeed, among the four objects common for the samples used in these two
studies, only for one (\einc) the masses determined by the two methods are
consistent within the uncertainties. For two objects (\pg\ and \rxj),
photometry-based mass estimates are about four times lower, while for the
least massive one, \sdss, masses differ by an order of magnitude. However,
the accretion disc sizes measured by \citet{morgan10} are still
several times larger than expected. Authors suppose that virial
mass estimates may still be systematically lower by a factor of $\sim
3$. 



The analysis method used by \citet{morgan10} was introduced by \citet{koch04}
and may be characterised as extensive light curve fitting. A number of
artificial amplification maps is generated, and the observational light curve
is compared with numerous simulated light curves with random parameters. This
technique was applied in a large number of studies such as
\citet{eigenbrodII} and \citet{hain12}. 
Since the technique requires multiple observational points, it
is perfect for Einstein's cross and other targets of monitoring programmes. 

The Monte Carlo microlensing analysis is rather resource-consuming. As an
alternative to computationally-extensive methods, it
is reasonable to study individual high-amplification events that
are most sensitive to the spatial properties of the source. Primarily,
high amplification events are associated to fold caustic crossings,
when a pair of new microlensing images appears or disappears, and the
point-source amplification diverges. 
Caustic crossings by standard discs are good models for some of the
observed amplification events \citep{GGG06,koptelova07}. It may be shown
that point-source amplification during a caustic crossing is divergent
for a disc with a small inner radius $r_{in}$ as $\mu \propto
r_{in}^{-1/4}$, hence brightness maxima in
microlensing curves are best fit for resolving the innermost parts of
the disc \citep{agolkrolik99}. Indeed, as it was shown by \citet{jaro}, lensing
curves differ for accretion discs of different inclinations and Kerr
parameters of the accretor. Amplification curves primarily differ
near their maxima. 

Caustic crossing events allow to study the structure of the innermost
parts of accretion discs where general relativity effects are
important. Two processes are expected to influence
considerably the observed light curves: light bending and Doppler
boosting (due to matter motion in the disc as well as due to frame
dragging). Both make brightness distributions asymmetric and strongly
dependent on the inclination angle. 
In this paper we aim on estimating the influence of relativistic
effects on the amplification curves created by straight caustic
crossing events. We also apply the results of our calculations to
three high amplification events and show that some of their features
are probably connected to relativistic effects. 

The paper is organized as follows: first we describe the archival
observational data we use. Simulation technique applying Kerr geodesic
calculation software is considered in section
~\ref{sec:method}. Fitting results for the  three putative
caustic crossings are given in section ~\ref{sec:res} and
discussed in section ~\ref{sec:disc}.

\section{Observational Data}\label{sec:obs}

The current number of lensed quasars where microlensing effects were found
is about several tens \citep{pooley07,morgan10,jimenez}. However, long
homogeneous observational series exist for few objects. The best
studied among these is the Einstein cross (\einc) that is a subject of
extensive photometric monitoring programs such as OGLE-II \citep{ogle}
and -III \citep{ogle3}. 
We analyse two
high-amplification events that took place for two different images in
the years 1999 and 2000. We also found a considerable amount of archival data on
another object, \sbs, and interpret variations of the relative
image amplifications as a manifestation of microlensing amplification. 
A brief summary of the observational data we use in this work is given
in table \ref{tab:obs}.

\begin{table*}\centering
\caption{Basic information about the objects and the observational
  amplification curves used for analysis. 
For the \einc\ image A event, the numbers of GLITP data points for two
reduction techniques, ISIS and PSF, are given separated by a slash. 
} 
\label{tab:obs}
\bigskip
\begin{tabular}{l|ccc}
  &  \sbs\  &  \einc (A)  &  \einc (C)\\

\hline

source redshift  &  1.855  &  \multicolumn{2}{c}{1.695}\\
lens redshift    &  0.72  &  \multicolumn{2}{c}{0.039}\\

\hline

time span (V), JD-2450000  & & 1400$\div$1650  &  1200$\div$1650\\ 

number of points (V) &  --  &  53/52 & 83 \\

\hline

time span (R), JD-2450000 & 1200$\div$3000  & 1450$\div$1510  & -- \\  

number of points (R) & 253  &  51/49 & -- \\

\end{tabular}
\end{table*}

\subsection{\sbs}\label{sec:obs:sbs}

The object is doubly imaged, with the average flux ratio for the two
images A and B  about 2 \citep{burud02}. The lens is a relatively
distant ($z\simeq 0.7$) late-type elliptical with the estimated velocity
dispersion of $\sigma \sim 200\kms$, steep mass profile,
central convergence of $\kappa \simeq 0.5$  and possible signatures of
interaction with its environment \citep{auger08}. 
 B is about three times closer to the lens centre, hence we
expect moderate microlensing optical depth for B and small
($\tau\lesssim 0.1$) for A. Hence we are inclined to interpret the
observed flux ratio variations as microlensing of the B image. 

\begin{figure*}
 \centering
\includegraphics[width=\textwidth]{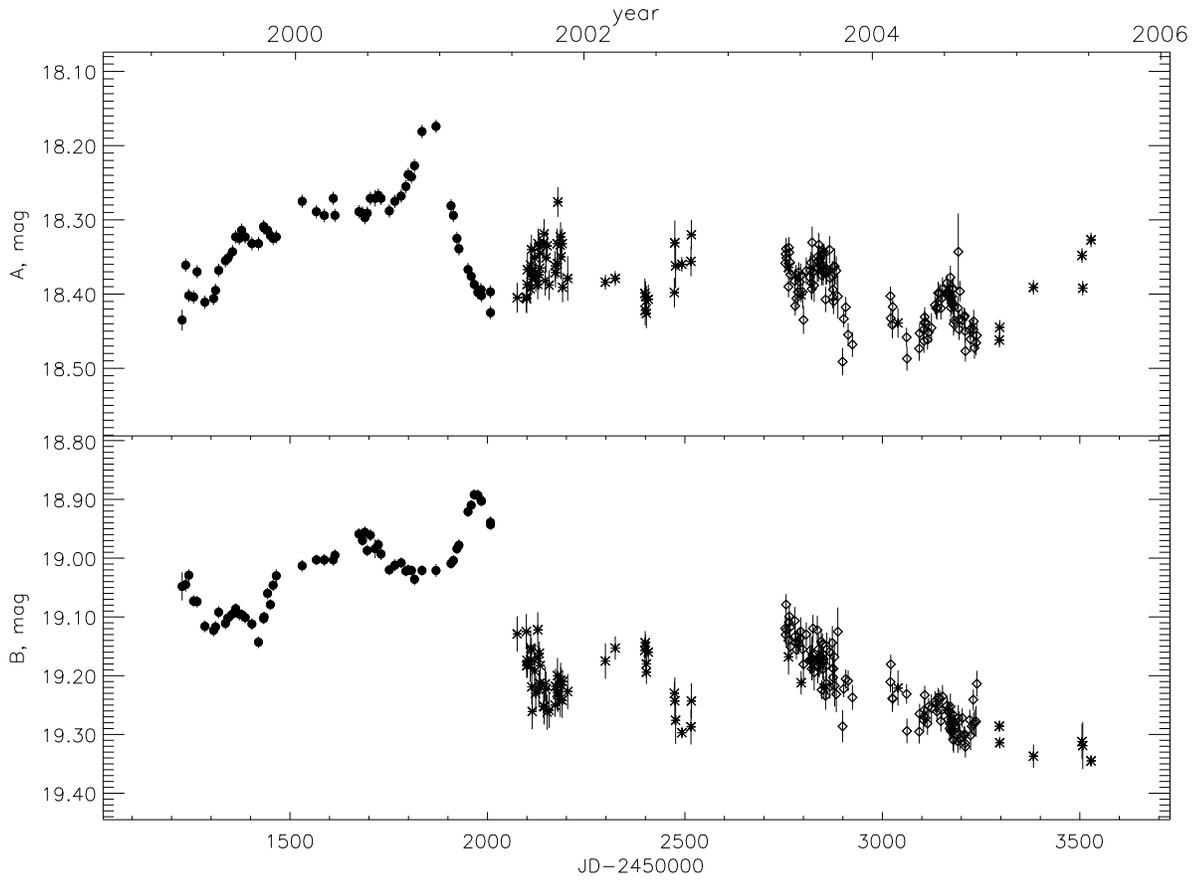}
\caption{R-band light curves for A (upper panel) and B (lower panel)
  images of \sbs. Filled circles, asterisks and diamonds represent
  NOT, RTT and Maidanak data (see text, section \ref{sec:obs:sbs}).
}
\label{fig:sbs:src3}
\end{figure*}

We used the three following sources of reduced photometric data
(R-band magnitudes for both images) covering a time span of about seven
years (see figure \ref{fig:sbs:src3}): 

\begin{itemize}
\item 58 data points obtained with the Nordic Optical Telescope (NOT)
 and  published by \citet{burud02}

\item 60 observations with the 1.5-m Russian-Turkish Telescope RTT
  \citep{khamitov06}, kindly provided by I.~Bikmaev and I.~Khamitov

\item 123 observations with the 1.5m AZT-22 telescope in
  Maidanak. These data were described, analysed and published by  
  \citet{sbs_maidanak}.

\end{itemize}

  All the photometry was performed in the standard optical R band that for the
source redshift of $1.855$ \citep{sbsz} corresponds to $\lambda \simeq
2000\div 2500\AAA$ in the source reference frame. 

Relatively large amount of data allows to improve the estimate of
delay time. Cross-correlating the time series for the two images we
find a broad peak at $\Delta t=127.6\pm 2.0$\,days
that we  interpret as the delay between the two images. Within the
uncertainties (that are hereafter calculated for 90\% significance
level), this value is consistent with the 130$\pm$3 day delay found by
\citet{burud02} and \citet{sbs_maidanak} and with the 128-day estimate
by \citet{khamitov06}. This value of $\Delta t$ is
subsequently used to shift the series of A and B fluxes and estimate flux
ratios. 

Merged time series consists of 241 unevenly-sampled R-band
observational points for each image (see figure \ref{fig:sbs:src3}).
 To minimize information losses but
to avoid unjustified interpolation (on periods of time longer than tens
of days), we construct flux ratios by linearly interpolating the
magnitudes between the observational points. The resulting flux ratio
series consists of points of two kinds: $(i)$ B image fluxes shifted
backwards by $\Delta t$ and divided by the interpolated values of A
image flux and $(ii)$ interpolated B image fluxes divided by the A
image fluxes shifted forward by $\Delta t$. We did not interpolate
over time gaps longer than 45d (about 16d in the frame co-moving with
the object), therefore the resulting time series contains only 360
points, not entirely independent (see also section
\ref{sec:res:sbs}). 

\begin{figure*}
 \centering
\includegraphics[width=\textwidth]{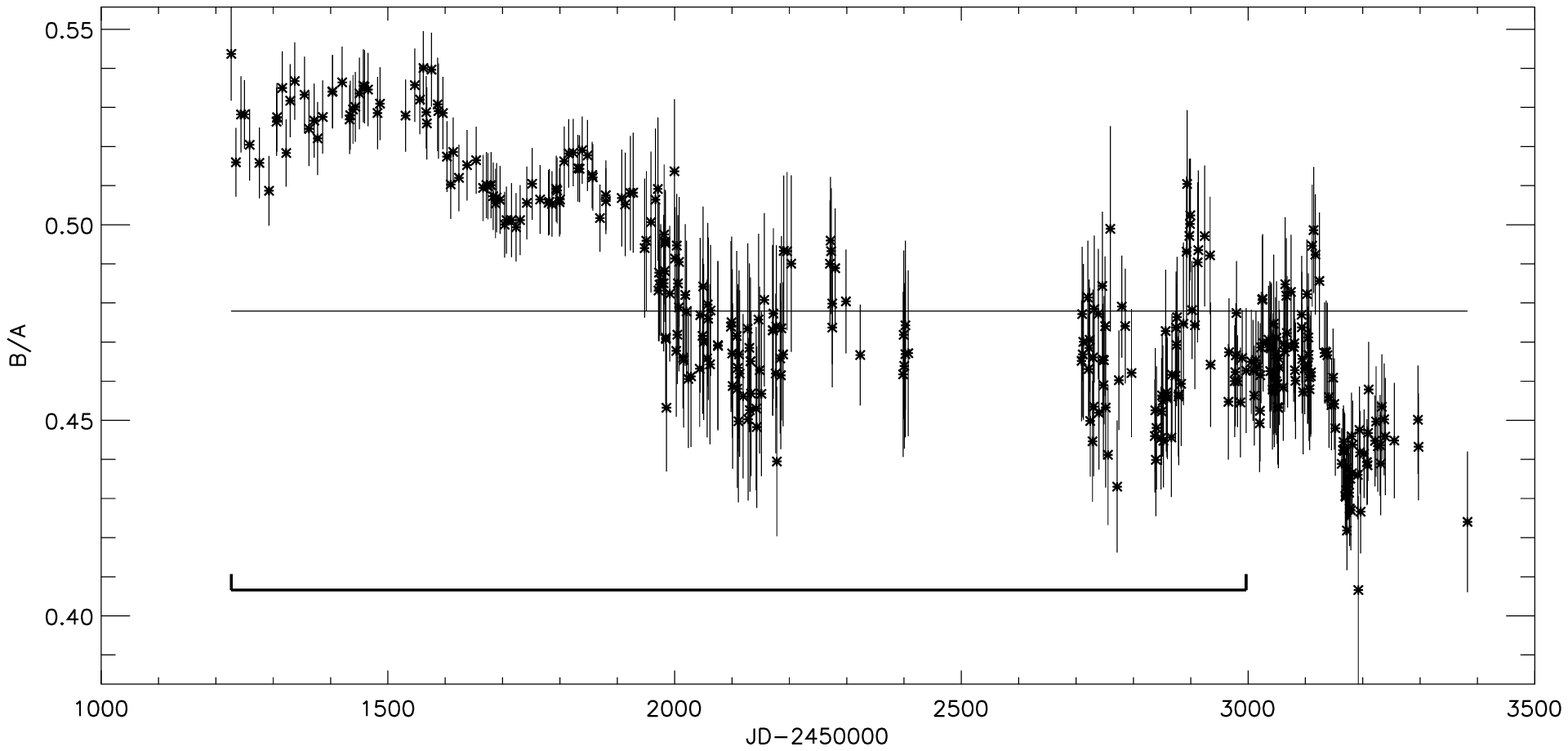}
\caption{Ratios of delay-corrected fluxes of images A and B of the
  lensed QSO \sbs. Thin horizontal line is the mean ratio value, thick
  horizontal line with ticks marks the time span used for our
  analysis. 
}
\label{fig:sbs:ratios}
\end{figure*}

Flux ratio curve (see figure \ref{fig:sbs:ratios}) demonstrates a
maximum at the Julian date $JD \sim
2451500$ and another one at $JD \sim 2451800$, that form together
an about two-year long structure similar to standard disc
amplification curves (see next section). For the remaining two-thirds
of the curve, $B/A$ ratio is stable and close to 0.47 with standard
deviation around four percent. 
From the resulting time series, we excluded the
final portion of the curve between $JD = 2453000\div 2453600 $ where the
data are sparse and the observed variability is difficult to relate
to the high-amplification event in the earlier part of
the data (see figure \ref{fig:sbscurve}). The resulting
amplification curve consists of 253 data points.

\subsection{\einc}\label{sec:obs:einc}

We use R- and V-band photometric data on the images A and C of
\einc. The data were taken from two sources:

\begin{itemize}

\item OGLE-II Huchra's lens monitoring program \citep{ogle}, the data are
available at
\url{http://ogle.astrouw.edu.pl/cont/4_main/len/huchra/huchra_ogle2.html}
and contain V-band magnitudes for the four images

\item GLITP archive \citep{glitp}: V- and R-band magnitudes for all the
  images, reduced data are available at
\url{http://wela.astro.ulg.ac.be/themes/extragal/gravlens/bibdat/engl/lc_2237.html}.

\end{itemize}

\begin{figure*}
 \centering
\includegraphics[width=\textwidth]{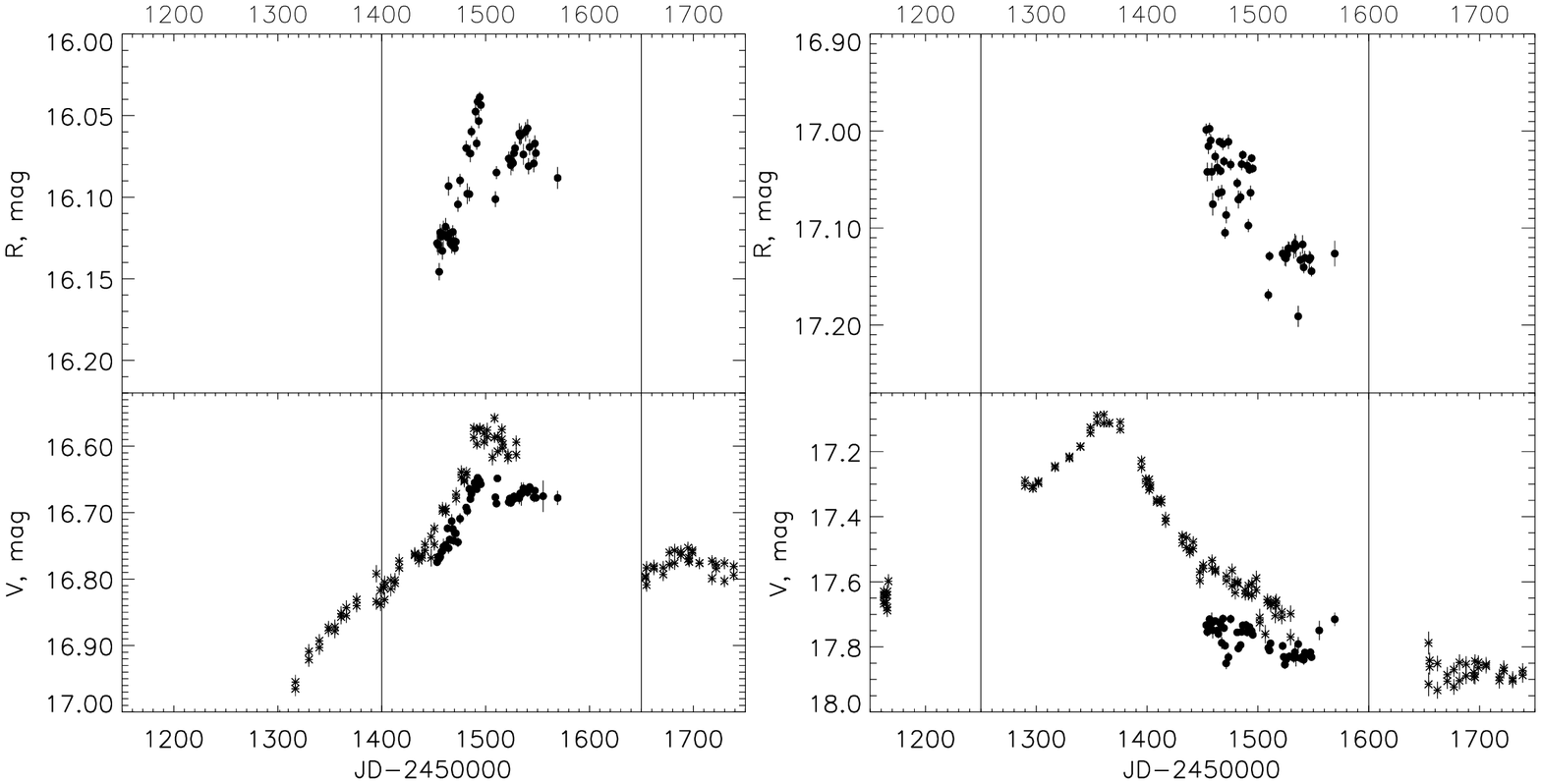}
\caption{Raw light curves of the images A (left) and C (right) of
  \einc\ used in the current work. GLITP(ISIS) data are shown by filled
  circles, OGLE data points by asterisks.
Date intervals used for fitting are restricted by vertical lines.}
\label{fig:einc:VR}
\end{figure*}

The light curves are shown in figure \ref{fig:einc:VR}. 
The data sample is similar to that used by
\citet{koptelova07}. 
We use GLITP data reduced by two different techniques: ISIS photometry (the
magnitudes are available at the web page given above) and PSF-fitting (the
magnitudes are meant to be available for download but the link is broken,
hence we asked Elena Shimanovskaya who has kindly provided us with these
magnitudes). The two reduction techniques are compared in
\citet{glitp}. Magnitudes reduced by different techniques are generally
consistent but deviate considerably (by about $0\magdot{.}03$) near the maximum
of the image A high-amplification event. Unfortunately, the maximum interferes
with a period of bad visibility of the object (January/February 2000). 
OGLE points show unreasonably large scatter near JD=2451500. Besides, consistency between OGLE
and GLITP data during this period is poor and insufficient for our purposes 
(see figure \ref{fig:einc:VR}, left panel). Better but still significant
deviations
between the GLITP magnitudes reduced with different methods (we show only
ISIS-reduced data to prevent the figure from overcrowding; the two GLITP  light
curves are shown together in figure \ref{fig:eincA}). 
Therefore we exclude OGLE data from subsequent analysis for the image A
event. 

GLITP data during the maximum of image A brightness demonstrate a
double-peaked structure that
we are inclined to interpret as a signature of the inner disc
structure. Overall behaviour near Julian dates 2451520$\div$2451540 suggests
that the dip is real and has an amplitude of about 0$\magdot{.}$02. 

For the C image event, most of GLITP observational points are far away from
the high amplification event maximum, and we finally fit only the V-band OGLE
data. 

There is strong evidence that the individual time
delays for the images of \einc\ are very small, such as several days 
(see \citet{koptelova_delays} and references therein). We can
make use of the relative proximity of the lens that
makes the microlensing variability of individual images much faster
and stronger 
than the intrinsic variability of the source. The studied high
amplification events last less than a
year, and variability of individual images does not show any detectable
correlation on these time scales. Einstein cross is known for its lack
of variability, the maximal gradients of intrinsic variability are
considerably smaller than microlensing trends (see for example
\citet{eigenbrodI}). 

\section{Amplification curve simulation technique}\label{sec:method}

\subsection{Basic assumptions}\label{sec:base}

All the three high-amplification events are good candidates for
caustic crossings. We model them by convolving a straight caustic crossing curve
with a standard disc profile integrated over one dimension. Simplified
flat and non-relativistic (section \ref{sec:plaindisc}) and general relativistic
(section \ref{sec:kerr}) disc models were used. In both cases we
consider a multi-blackbody disc with monochromatic intensity
obeying the Planck law with the temperature determined by the standard
accretion disc model:

\begin{equation}\label{e:idisc}
\begin{array}{l}
I \propto
\frac{1}{\exp\left(\frac{h\nu}{T(r)}\right)-1}
= \\
\qquad{}
=\frac{1}{\exp\left(\left(r/r_d\right)^{3/4}\left(1-\sqrt{r_{in}/r}\right)^{1/4}\right)-1},\\
\end{array}
\end{equation}
\noindent

where r is the radial coordinate, and $r_d$ and $r_{in}$ are,
correspondingly, the radial scale of the disc and its innermost stable
orbit radius defined as:

\begin{equation}\label{e:rd}
r_d = \left(\frac{k}{hc}\frac{\lambda}{1+z}\right)^{4/3} \left(\frac{3}{2}
\frac{l}{\eta} \frac{G^2M^2}{c \sigma \kappa }\right)^{1/3} 
\end{equation}

\begin{equation}\label{e:rin}
r_{in} = x_{\rm ISCO}(a) \frac{GM}{c^2} 
\end{equation}

Here, $k$, $\sigma$ and $h$ are Boltzmann, Stephan-Boltzmann and
Planck constants, $c$ is the speed of light, $l$ and $\eta$ are
Eddington ratio and overall accretion efficiency, $\varkappa$ is
Thomson opacity, $\varkappa \simeq 0.35\rm cm^2\,g^{-1}$ for Solar
composition. 
Dimensionless innermost stable circular orbit radius $x_{\rm
  ISCO}(a)$, as well as efficiency $\eta=\eta(a)$, may be found
analytically as functions of dimensionless Kerr parameter $a$
\citep{bardeen72}. 

Expression (\ref{e:idisc}) was derived for monochromatic intensity (no
matter $I_\nu$ or $I_\lambda$) but is to a high accuracy valid even
for broad-band photometry if most of the radiation comes in the form of
a smooth continuum. Mean effective wavelength varies by less than 2\% for
the spectral slope varying from $p=-1$ to $p=1$, where $F_\nu
\propto \nu^p$. The above definition of $r_d$ is identical to 
the characteristic radius used by \citet{morgan10}. Note however that
it is generally sufficiently smaller (by a factor of $\sim 2.44$ for
a standard disc with no inner edge) than the half-light radius. We
will see below that
for the shape of amplification curve, the quantity of principal
importance is the ratio of the two radial scales given by (\ref{e:rd}) and
(\ref{e:rin}):

\begin{equation}\label{e:rrat}
\begin{array}{l}
X = \frac{r_{d}}{r_{in}} =\\
\qquad{} =\left(\frac{k}{h}\frac{\lambda}{1+z}\right)^{4/3} \left(
\frac{3}{2}\frac{l}{\eta(a)} \frac{c}{GM\sigma\varkappa_T} \right)^{1/3}
\simeq \\
\simeq  92 \left( \frac{\lambda/1 \mu}{1+z}\right)^{4/3}
\left(\frac{l}{0.25}\right)^{1/3}
\left(\frac{\eta(a)}{0.1}\right)^{-1/3}
\left(\frac{M}{10^9\Msun}\right)^{-1/3} x_{ISCO}^{-1}(a)\\
\end{array}
\end{equation}

For fixed $X$, the differences between amplification curves are
important only for high inclinations and are connected to relativistic
effects, namely  Doppler boosting and light bending. Variations of the
Kerr parameter may change $X$ by a factor of
$\max(x_{ISCO})/\min(x_{ISCO}) =9$. Inversely, one may
reasonably estimate $X$ by fitting the amplification curve and arrive to
a tightly correlated pair of uncertain $a$ and $M$. Having
reasonable mass estimates, one may thus make an estimate for $a$, and
{\it vice versa}. 

Below, we fix  the Eddington ratio to $l = 0.25$. The black hole
masses are estimated as $M = 8.8\times 10^{8}\Msun$ for \sbs\ and
$M=9\times 10^{8}\Msun$ for
\einc\ by \citet{morgan10}. These mass
estimates were obtained using emission line profiles (see references given by
\citet{morgan10}) and virial
relations \citep{VP06}. Accuracy of these estimates is low, about
0.3dex. Our analysis yields generally higher masses poorly consistent
with the virial estimates (see below section \ref{sec:res}). 
In fitting the observational data, we first find $X$,
inclination $i$ and positional angle $\psi$ and then optimize for the
mass and Kerr parameter, fixing $X$. 

\bigskip

\begin{figure}
 \centering
\includegraphics[width=\columnwidth]{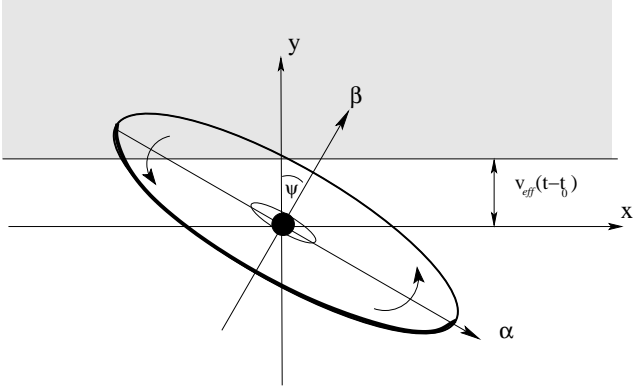}
\caption{A qualitative scheme showing the two coordinate systems in
  the picture plane. Shaded is the area behind the caustic where
  amplification is higher. Disc is inclined with respect to the
  observer, and its closer side is shown by a thicker curve. 
}
\label{fig:disksketch}
\end{figure}

Caustic is considered a straight line defined by condition $y=v_{eff}\times
(t-t_0)$ at the given time $t$. Let $x$ and $y$ be the coordinates in the
picture frame along and across the caustic, respectively. The origin
of this coordinate system coincides with the black hole centre. 
In the general case, the disc is
inclined, and it is convenient to use the coordinate system $(\alpha,
\beta)$ defined by the major and the minor axes of the disc projection
upon the picture frame (as in \citet{agoldexter}). The two coordinate systems in the picture
frame are connected by rotation by an angle of $\psi$ that has the
meaning of relative positional angle of the normal to the disc with
respect to the normal to the caustic (see a sketch in figure
\ref{fig:disksketch}).  Without any loss of generality we assume that
the relative motion of the source and lens is normal to the caustic
itself. Integration over one direction is convenient to consider as
projection upon the orthogonal direction.

Effective velocity $v_{eff}$ is calculated as the relative proper
motion multiplied by the angular size distance $D_S$ toward the
object. Our analysis is sensitive only to its component perpendicular to
the caustic direction. The velocity is connected to the peculiar
spatial velocities of
the QSO (\vector{v_S}) and the lensing galaxy (\vector{v_L}) in the
following way:

\begin{equation}\label{e:veff}
v_{eff} = \frac{\vector{v}_S \cdot \vector{n}}{1+z_S} -
\frac{\vector{v}_L \cdot \vector{n}}{1+z_L} \frac{D_S}{D_L}+\frac{\vector{v}_o \cdot \vector{n}}{1+z_L} \frac{D_{LS}}{D_L}
\end{equation}

Third term describes the peculiar motion of the observer, $v_o \sim 370\kms$
is known from the dipole constituent of cosmological CMB variations (see for
example \citet{lineweaver}). 
Unit vector $\vector{n}$ is normal to the caustic and lies in the
picture plane. 
Reasonable limit for peculiar motion velocities of individual
galaxies is $v \lesssim v_{max}\simeq 2000\kms$ that corresponds to
about $2\div3$~rms values for the radial peculiar velocity
distribution \citep{RS96}. This limit may be converted to the
condition for the possible effective transverse velocity $v_{eff}
\lesssim v_{max} \sqrt{(1+z_S)^{-2}+(1+z_L)^{-2} \times
(D_S/D_L)^2}$. For \sbs, \vector{v_S} and \vector{v_L} contributions
are of the same order because $D_S / D_L \simeq 1.2$. 
In this case, $v_{eff} \lesssim
1600\kms$. For \einc, the lens is about ten times closer ($D_S /D_L \simeq
11$) and $v_{eff} \lesssim 20\,000\kms$.



\subsection{Simplified standard disc}\label{sec:plaindisc}

Besides the more sophisticated model described below, we use a simple
disc approximation ignoring all relativistic effects.
In this approximation and for the straight caustic case, inclination does
not affect the observed shape of the amplification curve. 
This is valid for a thin disc of arbitrary inclination as long as the disc is
thin enough ($h/R \ll \cos i$, where $i$ is inclination). 

We are interested in intensities integrated over one dimension. Here,
we provide the general form for the integral over one direction (we
make a substitute $t=x/y$): 

\begin{equation}\label{e:discplane}
\begin{array}{l}
I_1(y)=\int I(x,y) dx \propto \\ 
\qquad{} \propto \sqrt{\frac{J}{K}} y
\int_{t_{in}}^{+\infty} \frac{dt}{\exp\left(\left(\frac{y^2}{r_d^2} J\times (1+t^2) \right)^{3/8} f^{-1/4}\right)-1}, 
\end{array}
\end{equation}

where:

\begin{equation}\label{e:discplane:K}
K = K(i, \psi)=\cos^2\psi+\frac{\sin^2\psi}{\cos^2i},
\end{equation}

\begin{equation}\label{e:discplane:J}
J=J(i, \psi)=\sin^2\psi+\frac{\cos^2\psi}{\cos^2i} -\frac{\sin^2\psi
  \cos^2 \psi}{K(i, \psi)} \tan^4i,
\end{equation}

\begin{equation}\label{e:discplane:f}
f=1-\left(\frac{1}{J}\frac{r_{in}^2}{y^2} \frac{1}{1+t^2}\right)^{1/4}
\end{equation}

\begin{equation}\label{e:discplane:tin}
t_{in} =\left\{ 
\begin{array}{lc}
0 & \mbox{ if }y\ge r_{in}/\sqrt{J}\\
\sqrt{1-\frac{1}{J}\left(\frac{r_{in}}{y}\right)^2} & \mbox{ if }y< r_{in}/\sqrt{J}\\
\end{array}
\right.
\end{equation}

If the influence of the inner radius is negligible, the
one-dimensional intensity profile is identical to
the face-on disc intensity profile with the radial scale of
$r_d^\prime = r_d / \sqrt{J}$. Even if the
correction factor $f$ is taken into account,
profile shapes do not depend on the angles $\psi$ and $i$, because the
integral (\ref{e:discplane}) contains the angles and coordinates only
in combinations
$y\sqrt{J}/r_d$ and $y\sqrt{J}/r_{in}$ that affects only the stretch
factor of the curve for given $X$. 



\begin{figure}
 \centering
\includegraphics[width=\columnwidth]{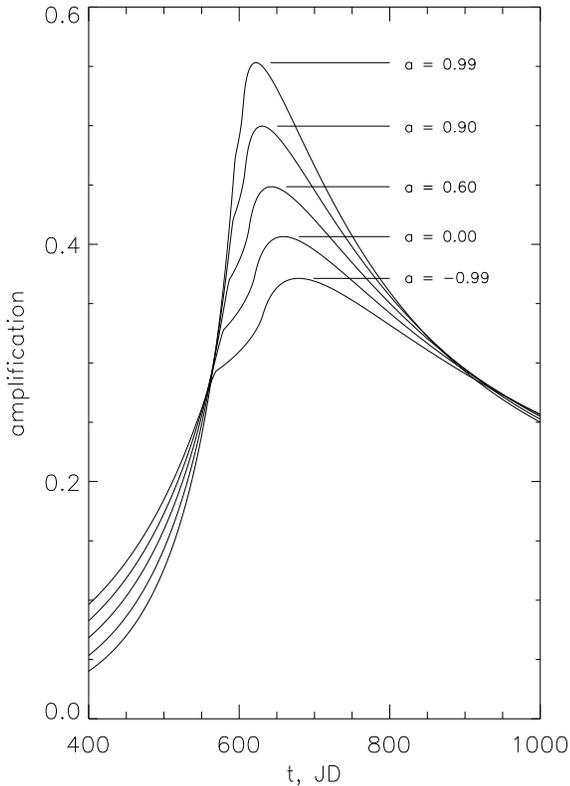}
\caption{Amplification curves for a simplified disc with a fixed mass
  ($M=10^9\Msun$ at comoving $\lambda=2300\AAA$) of the black hole and variable Kerr
  parameter. 
}
\label{fig:simple:shape}
\end{figure}

The influence of the inner disc edge is included by
the single parameter $X = r_d / r_{in} $ (see previous subsection) determining
the shape of the amplification curve. In figure \ref{fig:simple:shape}
the dependence of the amplification curve on this parameter is shown
for a fixed black hole mass. Kerr parameter is varied from -0.99 to
0.99. 

\subsection{Kerr black hole disc of arbitrary inclination}\label{sec:kerr}

Photons move along null geodesics that we calculate using {\tt
  geokerr} code \citep{agoldexter}. Disc was ascribed a constant
relative thickness of $h/r = 0.01$ that does not affect the results
much as long as $\cos i > h/r$. The maximal inclination value we
use is $i=88^\circ$ that implies $\cos i \simeq 0.03$. We linearly
interpolate the radial coordinate to the point where $\cos i = \pm h/r$ to
determine the intrinsic intensity of the locus in the disc
corresponding to
current values of $\alpha$ and $\beta$. For this point, using the conserved
angular momentum $k_\varphi$ and the energy at infinity $k_t$ of the
photon, we connect the observed photon energy (equal to $k_t$) to its
energy in the frame co-rotating with the disc. Doppler-factor is
calculated as:

\begin{equation}\label{e:delta}
\delta = \frac{\nu_{obs}}{\nu_{em}} = \frac{k_t}{k_i u^i} = \frac{1}{u^t \left( 1 + \Omega l\right)}
\end{equation}

Here, $u^t= 1/\sqrt{-(g_{tt} + 2g_{t\varphi}\Omega +
  g_{\varphi\varphi}\Omega^2)}$ is the null component of the disc
rotation four-velocity (where metric signature is -+++), $l=k_\varphi/k_t=\alpha
\sin i$ is the net angular momentum  of the photon. 
For co-rotating discs, 

\begin{equation}\label{e:omega}
\Omega = \frac{1}{r^{3/2}+a},
\end{equation}

in $c^3/GM$ units, while the radius $r$ is in
$GM/c^2$. Counter-rotating discs may be also considered in this
formalism by changing the sign of $a$. Hereafter we will use negative
$a$ values to describe the case of counter-rotation. 

Observed intensity is integrated over a fixed (observer-frame)
wavelength range and is therefore approximately proportional to the
monochromatic intensity $I$ at the central observer-frame
frequency $\nu_{obs}$. For a local blackbody-like spectrum (we
consider $I_\nu$ for convenience):

\begin{equation}\label{e:irel}
\begin{array}{l}
I_\nu \simeq \Delta\nu_{obs} \delta^3 I^0_\nu(\nu_{em}) \propto \\
\qquad{}
\propto \left(\exp\left(\frac{1}{\delta}\left(r/r_d\right)^{3/4}\left(1-\sqrt{r_{in}/r}\right)^{1/4}\right)-1\right)^{-1}\\
\end{array}
\end{equation}

Doppler boost thus contributes only through direct frequency
shift. At large distances ($r \gg r_{in}$ and $r \gg r_{d}$) its effect on the local
intensity is still important, primarily due to the rapid fall-off
of the Planck law with frequency. 
For inclined discs, intensity distribution along
the major axis is asymmetric even at large distances from the black
hole. Up to the two leading terms in $1/r$, $1 / \delta \simeq 1+
\sin(i) / \sqrt{r}$, and relative intensity difference between the
approaching ($I^+$) and receding ($I^-$) sides of an inclined disc is:

\begin{equation}\label{e:idiff}
\frac{I^+-I^-}{I^++I^-} \simeq \left( r/r_d\right)^{3/4}
\frac{\sin(i)}{ \sqrt{r}} \propto r^{1/4}
\end{equation}

Above estimate was made in the assumption that the asymmetry
is small. The actual value of asymmetry is evidently limited by the
value of $\max\left( (I^+-I^-) / (I^+ + I^-)\right) =1 $ 
when one side of the disc is much brighter than the other. 
The approaching side of an accretion disc is generally about two times
brighter than the receding at $r\sim 10\div 100GM/c^2$. 
Weighed disc centre at large
distances is effectively shifted by $\Delta r \simeq  \sin i \times  r^{1/2}$.

\subsection{Calculation of amplification curves}

For calculating the shapes of null geodesics, we use the public code
of \citet{agoldexter}. Then we take into account Doppler shifts and
Doppler boosts in the way described above. For given Kerr parameter
$a$, inclination $i$ and relative positional angle $\psi$, 
intensity is calculated on a rectangular grid log-uniform in $x$ and
$y$ coordinates. 
Then the two-dimensional
brightness distribution is integrated over one dimension:

$$
I_1(y)=\int_{-\infty}^{+\infty} I(x,y) dx
$$

Then, the microlensing amplification curve for the straight caustic
case may be calculated in the following way (see for example
\citet{koptelova07} who used a similar approach):

\begin{equation}\label{e:ampf}
\mu(t) = \mu_1 \sqrt{\zeta_0} \times \frac{\int
  I_1(y) \times \Delta y^{-1/2}\times \Theta(\Delta y)dy}{\int I_1(y)dy} + \mu_0
\end{equation}

Here, $\Delta y = y-v(t-t_0)$, and $\Theta(x)$ is Heaviside function:
$\Theta(x>0)\equiv 1$ and $\Theta(x\leq 0)\equiv 0$. Spatial scale
factor $\zeta_0$ is chosen according to \citet{WKR93}:

\begin{equation}\label{E:zeta0}
\zeta_0 = \sqrt{\frac{4G\Msun}{c^2} |1-\kappa_c|\frac{D_{LS}D_S}{D_L}}
= r_{Ein}(\Msun)\times\sqrt{1-\kappa_c} 
\end{equation}

Here, $\kappa_c$ is the continuous contribution to the total
convergence. If we adopt $\kappa_c=0.5$, for
\sbs\ and \einc, the spatial scale is $\zeta_0 \simeq 3\times 10^{16}$
and $1.3\times 10^{17}\rm cm$, respectively. 

Effective transverse velocity $v$ and the fold-crossing epoch $t_0$ are
considered as free parameters. For every model (every given one-dimensional
brightness distribution), $v$ and $t_0$ are optimized for by
minimising $\chi^2$ for the observational amplification
curve. For the A image of \einc, two amplification curves were used for
the two filters, R and V, and we fit both
simultaneously. Our fitting has two additional parameters, $\mu_0$ and
$\mu_1$, that we calculate using linear regression for each iteration
of the optimisation process. In the case of \einc, two 
$\mu_{0,1}$ pairs were used for the two photometric bands. 
The ratio of the two coefficients, $\mu_1
/ \mu_0$, has the meaning of caustic strength and depends only on the
mass distribution in the lens \citep{KW89}.

\begin{figure}
 \centering
\includegraphics[width=\columnwidth]{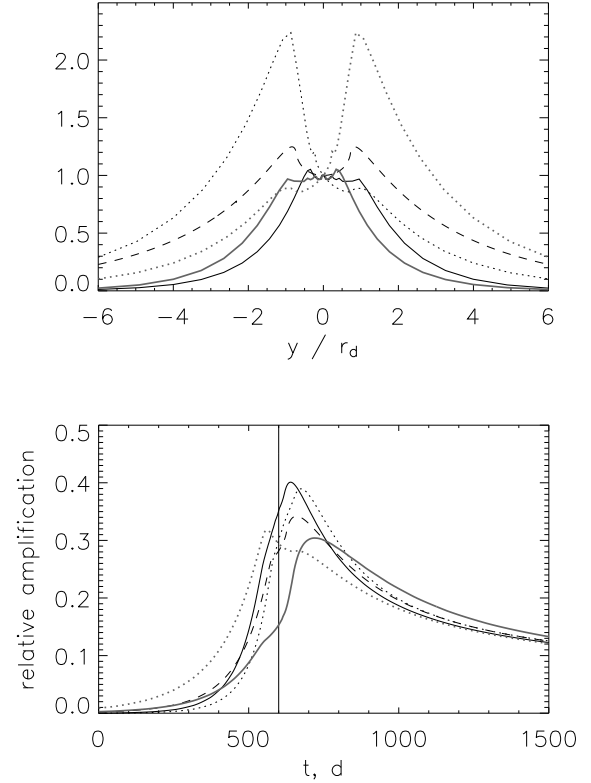}
\caption{One-dimensional brightness distributions (upper panel) and
  amplification curves (lower panel)
  for a disc inclined by $60^\circ$ around a $a=0.6$
black hole. Dashed curve is the simplified disc model,
black and grey curves correspond to caustic transition along the minor
and major axes of the inclined disc (solid and dotted curves differ in
propagation direction). 
}
\label{fig:xirel}
\end{figure}

\begin{figure*}
 \centering
\includegraphics[width=0.9\textwidth]{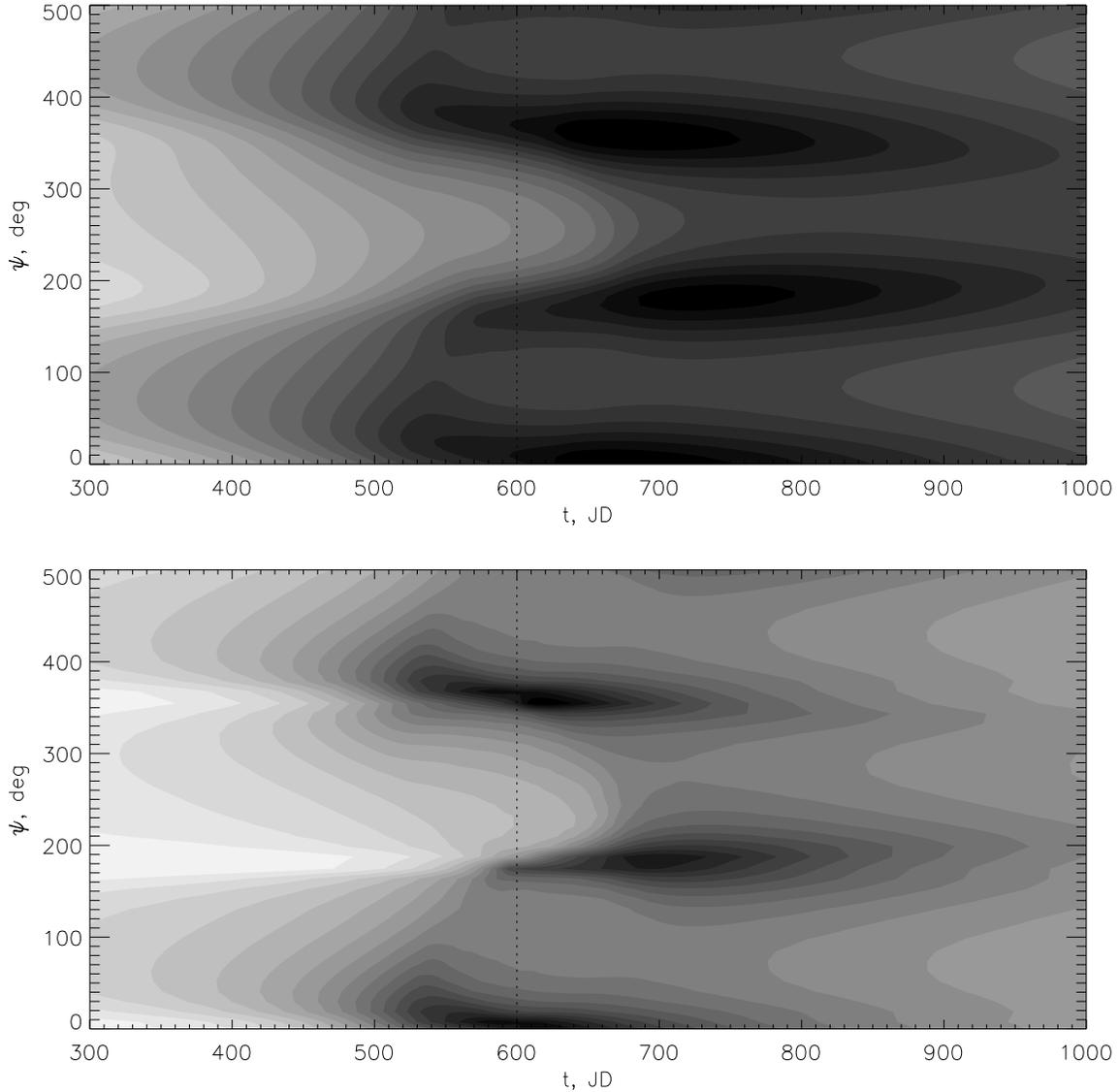}
\caption{ Amplification curves for an accretion disc around a black
  hole with $a=0.6$ inclined by $i=60^\circ$ (upper panel) and
  $i=85^\circ$ (lower panel) for variable positional
  angle $\psi$. Every horizontal slice corresponds to one curve. Scale
  is linear, maximal shade corresponds to maximal brightness.
}
\label{fig:animkerr}
\end{figure*}

Evidently, for the shape of the caustic-crossing amplification
curve, the effects of general relativity are of primary importance.
In figure \ref{fig:xirel}, we show four light curves emerging from
caustic crossings by a disc inclined by $60^\circ$ around a $a=0.6$
black hole, traversing a straight fold caustic in four directions:
along the minor axis ($\psi=0$ and
$180^\circ$) and along the major ($\psi=90$ and
$270^\circ$). Effective transverse velocity and black hole mass used
for the figures \ref{fig:xirel}
and \ref{fig:animkerr} were $5000\kms$ and $10^9\Msun$, respectively. 
Due to their asymmetry, inclined discs produce
amplification curves strongly
dependent on the relative positional angle. Figure
\ref{fig:xirel} demonstrates all the main complications arising from
relativistic effects. If the brighter part of the disc is amplified
while its dimmer part is still unaffected by the caustic,
amplification curve may become two-peaked with the relative intensity
of the two peaks strongly dependent on inclination, positional
angle and black hole rotation parameter (figure \ref{fig:animkerr}). 
An accretion disc surrounding a rapidly rotating Kerr black hole
observed at high inclination has a  bright compact ``hot spot'' that
influences strongly the amplification curve. 
Relativistic effects are capable, in particular, to qualitatively
explain the observed fine structure of the amplification events under
consideration.

\begin{figure*}
 \centering
\includegraphics[width=\textwidth]{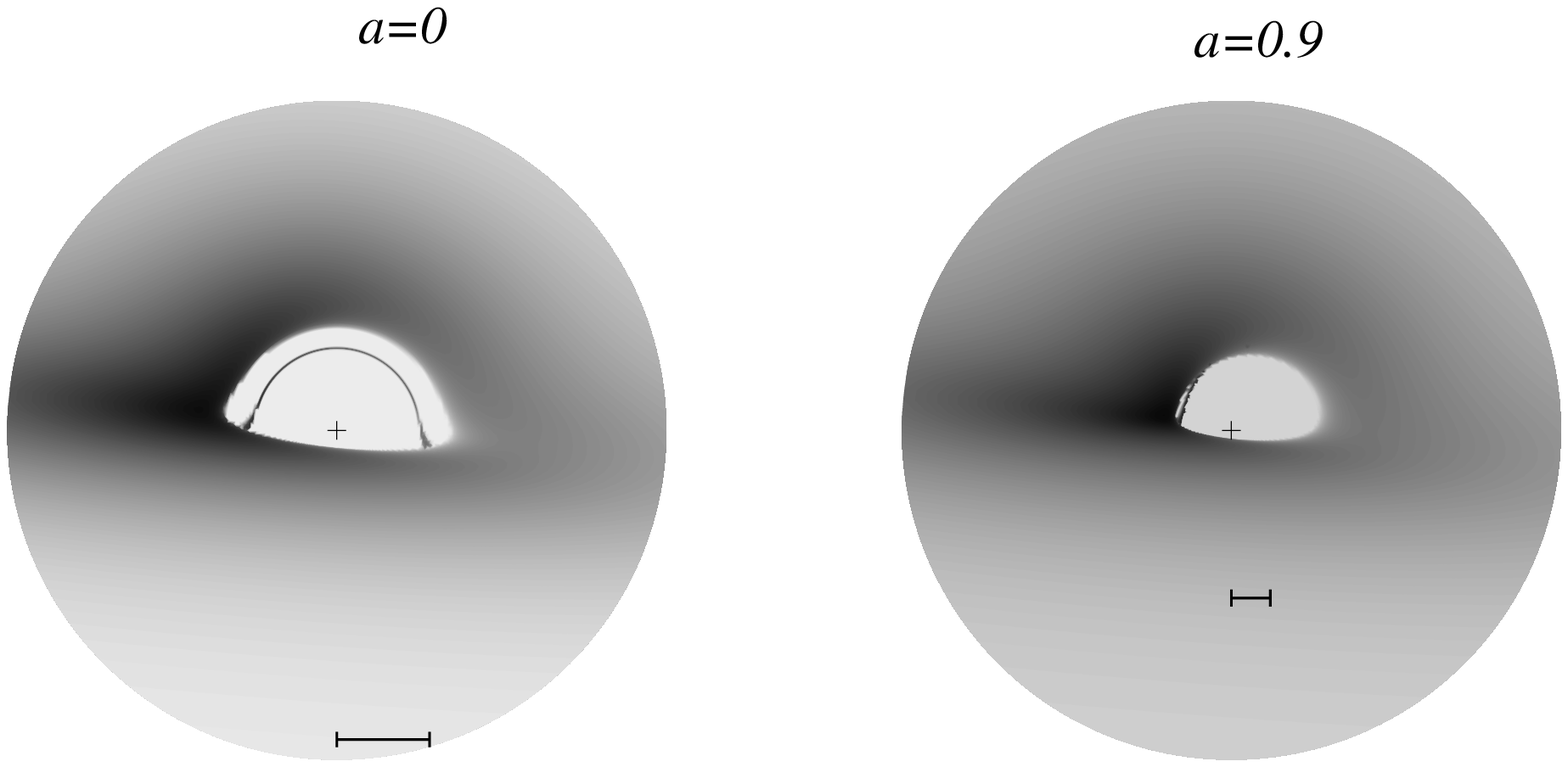}
\caption{Black hole shadows for $i=80^\circ$ and two Kerr parameters,
  $a=0$ and $0.9$. Secondary image is clearly seen for Schwarzschild
  black hole and is visible as an inclined nearly straight line at the
  approaching (left) side in the other case. A bar below the black
  hole has the length equal to the innermost stable orbit radius (6 and
  $\sim$2.3 in $GM/c^2$ units, respectively).
}
\label{fig:shadows}
\end{figure*}

 Test runs reveal a small but considerable contribution from the
photons belonging to the higher-order images produced by extreme
light bending near the black hole \citep{BD05}. Their contribution to
the total flux may be as high as several percent,
depending on the inclination and on the Kerr parameter. The figure of
10\% given by \citet{BD05} is an over-estimate for an optically thick
disc because at high
inclinations and high Kerr parameters considerable part of the black
hole is covered by the disc. 
Simulated black hole shadows with one higher-order image
visible are shown in figure \ref{fig:shadows}. Higher orders have
significantly smaller fluxes due to gradually decreasing solid angle. 
Flux of $n$-th image loop
decreases proportionally to its radial size $\Delta r$ (intensity is
approximately conserved). During caustic crossing, the additional
amplification factor is
proportional to $\Delta r^{-1/2}$, where the observed
projected width of the approximately annular image is $\propto \Delta r$. The resulting
contribution of the $n$-th order image scales as $\propto \Delta r^{1/2}\propto
\exp(-\pi n/b )$, where $b\simeq 2.7$ (in the so-called strong
deflection limit, see \citet{bozza}). Higher-order images are
thus exponentially damped. 

\begin{figure}
 \centering
\includegraphics[width=\columnwidth]{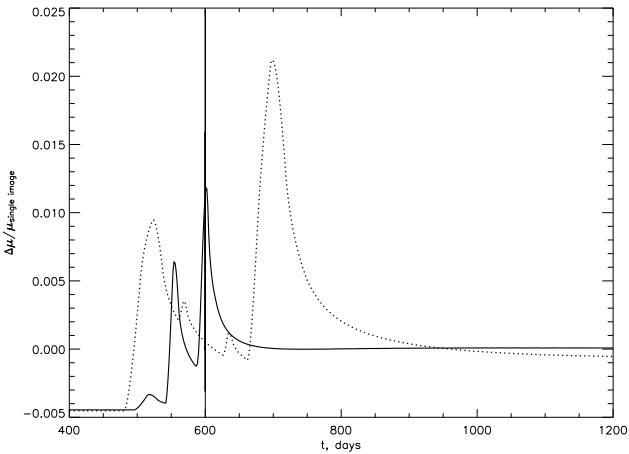}
\caption{Relative difference between the microlensing curves
  calculated in the single-image assumption and taking into account
  all the resolved secondary loops.
We set $M=10^{10}\Msun$, $a=0$, $i=80^\circ$ and $\psi=0$ (solid line)
and $90^\circ$ (dotted line). Effective transverse velocity is
$v_{eff}=1000\kms$, vertical line marks the instance of caustic
crossing by the centre of the black hole.
}
\label{fig:second_im}
\end{figure}

An example of higher-order contribution is given in figure
\ref{fig:second_im}. For high inclinations, it is generally about
a couple percent, but are less than one percent for $i\sim 0 $ and $a
>0 $. For counter-rotating black holes, the size of the inner hole is
the largest and the contribution of the secondary images reaches
2$\div$3 percent for characteristic QSO black hole masses $\sim
10^9\Msun$. The effect is also important for larger black
holes and at shorter wavelengths.  

\section{Results}\label{sec:res}

Fitting results with the simplified face-on disc and fully
relativistic disc models are given in tables \ref{tab:res:plain} and
 \ref{tab:res:kerr}, respectively.

\subsection{\sbs}\label{sec:res:sbs}

\begin{figure*}
 \centering
\includegraphics[width=0.9\textwidth]{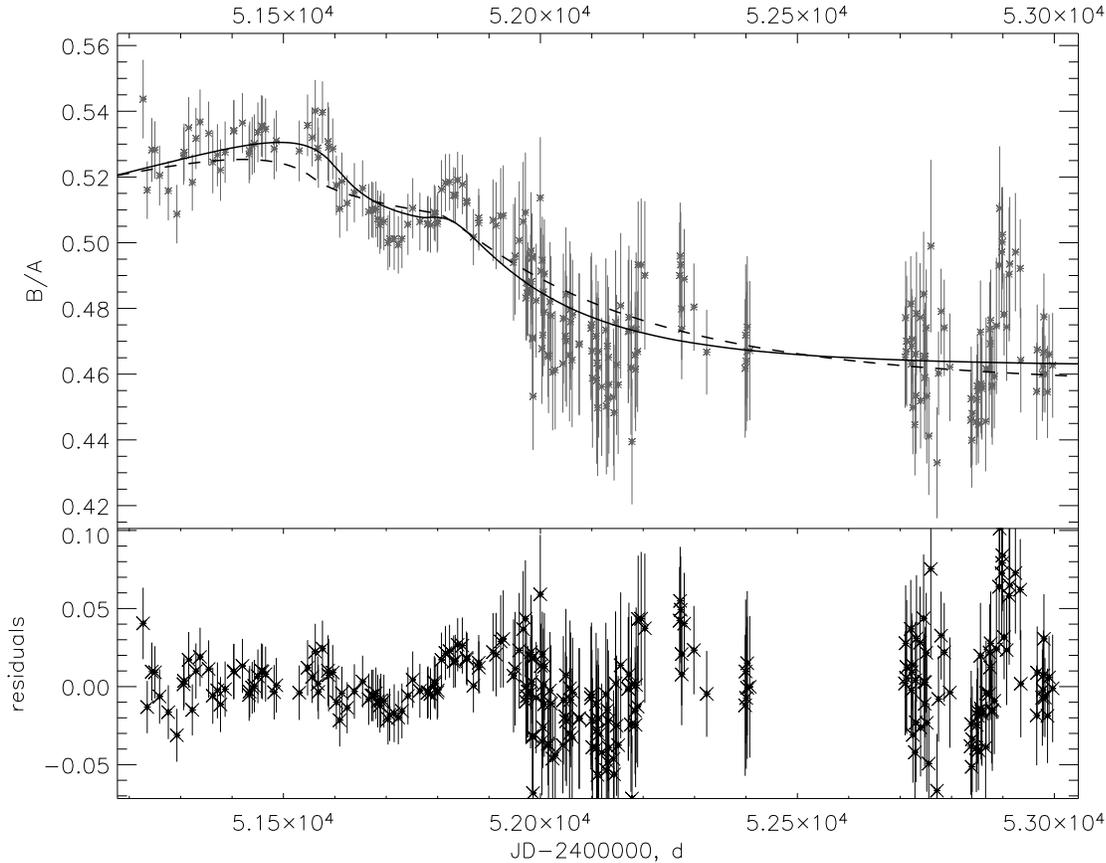}
\caption{R-band B/A flux ratio for \sbs\
  fitted with a Schwarzschild face-on ``plain disc'' without
  relativistic effects (dashed curve) and a Kerr multi-blackbody disc
  of arbitrary inclination and spin (solid).  The parameters of the best-fit
  Kerr disc model are: $a=-0.5$, $M=2\times 10^9\Msun$, $i=80^\circ$,
  $\psi=330^\circ$. Lower panel shows
  residuals with respect to the Kerr disc model. 
}
\label{fig:sbscurve}
\end{figure*}

\begin{figure*}
 \centering
\includegraphics[width=0.9\textwidth]{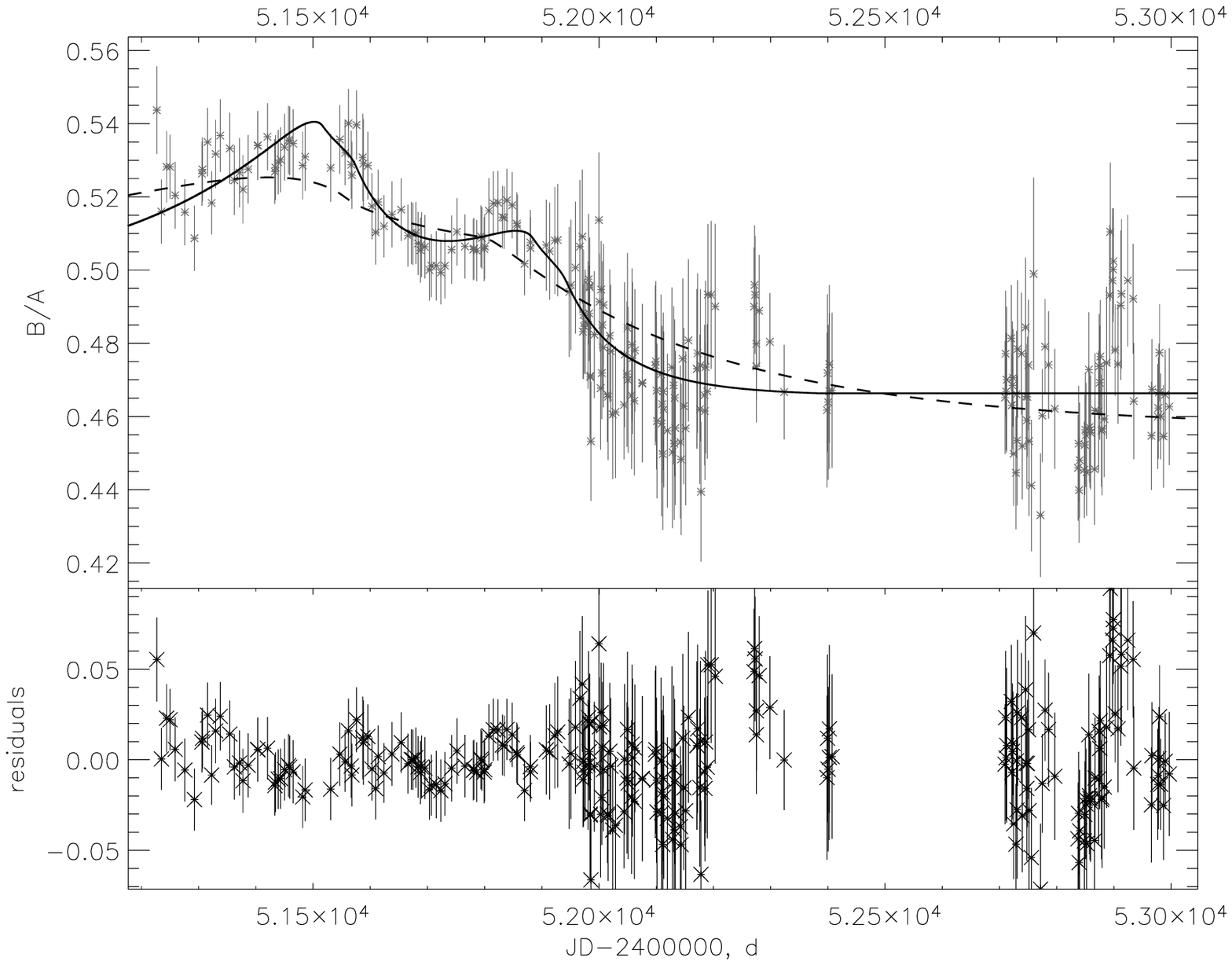}
\caption{R-band B/A flux ratio for \sbs\ fitted with a
  double-fold model (the best-fit curve parameters are given in the text, section
  \ref{sec:res:sbs}). Simplified disc single-fold model curve is shown
  for comparison  by a dashed curve.
}
\label{fig:twofold}
\end{figure*}

Fitting with the plain face-on disc model allows to approximately recover the
size of the inner hole. 
 Number of degrees of freedom is $253-5=248$ here (``optimistic''
 number, see below). We fixed the mass of
the black hole to $3\times 10^9\Msun$ and localized an absolute
minimum near $a=-0.4$. Corresponding $X$ value is $1\div 2$. 
The minimum is consistent with physical $-1 < a <
1$ only for the black hole masses in the range $(1.6\div 3.5) \times 10^9\Msun$. The
best-fit amplification curve is also shown in figures
\ref{fig:sbscurve} and \ref{fig:twofold} by a
dashed line. Here we limited the effective velocity by the value of
$v_{eff,max} = 2000\kms$. If we relax this limitation, the fit may be
improved (up to $\chi^2/DOF \sim 190$) by increasing the mass of the
black hole and the transverse velocity.

%
%

A total number of 708 fully relativistic models was calculated covering the
possible ranges of
inclination $0<i<90\deg$, relative position angle $0<\psi<360\deg$
and Kerr parameter for co-rotating ($1>a>0$) and counter-rotating ($-1<a<0$) cases. 
For every model, we make an optimization run finding the best-fit
$v_{eff}$ and $t_0$ and two amplification parameters $\mu_0$ and
$\mu_1$. Thus the model one-dimensional brightness distribution fixes
only the shape of the amplification curve that may be then stretched
and shifted in time ($t_0$ and $v_{eff}$) and in amplification factors
($\mu_{0,1}$). For Kerr disc models, we first fix the mass of the
black hole to $0.88\times 10^9\Msun$ 
and then, after finding the global minimum at
$a=-0.9^{+0.4}_{-0.05}$, $i=80\pm 5\deg$ and
$\psi=340\pm 4\deg$, fix $X$, $i$ and $\psi$ and vary the
mass and rotation parameter. 

In tables \ref{tab:res:plain} and \ref{tab:res:kerr}, 
we give the best fit parameters for the
amplification curve of \sbs\ fitted with the simplified disc model and with
a Kerr disc with the best-fit parameters. Fully relativistic disc
provides much better fit (see below this section). However, the observed
details in the amplification curve are even sharper and more
profound. This is a possible signature of a non-trivial structure of the
inner parts of the disc resulting from either disc tilts and warps
(see below section \ref{sec:disc:tilt}) or from additional energy
input from black hole rotation \citep{agolkrolik}. 

The apparently low values of $\chi^2$ should be addressed
separately. The procedure that we used to reproduce the amplification
curve (see section \ref{sec:obs}) minimizes the loss of observational
information but generates two sets of flux ratios (A image
fluxes interpolated over the observational epochs for B and {\it vice
versa}) that are not entirely independent. Therefore, the effective number
of degrees of freedom (DOF) is not the number of flux ratio data points ($N_F=$253) minus the
number of parameters ($N_P=5$ for simplified disc and $N_P=8$ for Kerr
metric) but less. Therefore in the tables we give two
estimates for the number of DOF: one optimistic $DOF_1 = N_F-N_P$,
and one pessimistic $DOF_2 = N_F/2-N_P\simeq 120$. The real
value is somewhere in between. For the simplified disc model, the
number of parameters is less by 3 ($i$ and $\psi$ lacking, and $X$
instead of $a$ and $M$).

For both DOF estimates, the difference in $\chi^2$ is significant:
F-test gives the probability of $7\times 10^{-10}$ (optimistic) and
$4\times 10^{-5}$  (pessimistic DOF estimate) for the more sophisticated model
to provide a better fit accidentally. The measured best-fit
parameters, primarily the high inclination, are
suspicious but not entirely unphysical. We propose that the real disc
should be inclined with respect to the black hole plane and its inner
parts should be thus strongly warped and distorted. In more detail we
discuss this issue in section \ref{sec:disc:tilt}.

\begin{table*}\centering
\caption{Results of amplification curve fitting with the simplified
  accretion disc model. Effective velocity is normalised by the disc
  radial scale (R-band for \sbs, V-band for \einc) in $10^{15}\rm cm $ units.}
\label{tab:res:plain}
\bigskip
\begin{tabular}{lccccc}
 & $r_d/r_{in} $  & $v_{eff}/r_{d,15}$,\kms  & $t_0$, JD-2450000d & $\mu_1/\mu_0$  & $\chi^2 /
  DOF$  \\
\hline
\sbs\ & 1.5$^{+0.6}_{-0.4}$ & 740$^{+80}_{-60}$  & 1676.3$_{-0.1}^{+4.7}$ & 1.05$\pm$0.17
& 209/246(120) \\
\hline
\multicolumn{2}{l}{\einc\ (A):} \\ 
ISIS &
$
\begin{array}{l} 
2.0\pm 0.03\mbox{ (R)}\\
1.6\pm 0.02\mbox{(V)}\\
\end{array}
$
& -2900$\pm 100$ &  1460$\pm 5$  & 
$
\begin{array}{l} 
2.4\pm 0.3\mbox{(R)}  \\
3.9\pm 0.4 \mbox{(V)}\\
\end{array}
$
 & 560/99 \\ 
PSF &
$
\begin{array}{l} 
2.6^{+0.13}_{-0.5}\mbox{ (R)}\\
2.12^{+0.1}_{-0.4}\mbox{(V)}\\
\end{array}
$
& -2920$\pm 120$ &  1455.0$\pm 2$  & 
$
\begin{array}{l} 
4.5^{+0.06}_{-0.2}\mbox{(R)}  \\
6.0^{+0.09}_{-0.4} \mbox{(V)}\\
\end{array}
$
 & 122/96 \\ 
\hline
\einc\ (C) & $\gtrsim$4.4 &
6100$_{-300}^{+400}$  & 1389.0$^{+0.2}_{-0.1}$ & 0.66$\pm$0.02
 & 183/77  \\ 
\hline
\end{tabular}
\end{table*}

\begin{table*}\centering
\caption{Results of amplification curve fitting with an inclined Kerr
  disc model. 
}
\label{tab:res:kerr}
\bigskip 
\begin{tabular}{ccccccccc}
 $r_d/r_{in} $  & $a$ & $M$  & $i$  & $\psi$  & $v_{eff}$  & $t_0$  & $\mu_1/\mu_0$  & $\chi^2 /  DOF$  \\
   & & $10^9\Msun$  & deg  & deg  & \kms &  JD-2450000\rm d  &  &   \\
\hline
\multicolumn{2}{l}{\bf \sbs:} \\
 2.15$\pm$0.05  &  -0.6$\div$0.2 &  1.7$\div$3.4 &
80$\pm$10  & 335$\pm$5 & 1500$\div$2400 & 1678$\pm$5 & 0.61$\pm$0.05 &  176/244(118) \\
\hline
\multicolumn{2}{l}{\bf \einc:}\\

\multicolumn{2}{l}{A image, ISIS} \\

1.7$\pm$0.1 (R) &  0.2$^{+0.1}_{-0.2}$  &  7$^{+2}_{-1}$  &
71$^{+15}_{-5}$ & 96$\pm $5  &   $-(3\pm1)\times 10^4$   &  $1509.9\pm 1.4$ &   0.91$\pm$0.06 (R) & 333/96  \\ 
  1.4$\pm$0.1 (V) &  &  &  &  &  &  &   0.92$\pm$0.03 (V) &  \\ 
\hline
\multicolumn{2}{l}{A image, PSF} \\

1.7$^{+0.6}_{-0.02}$ (R) &  -0.3$\div$0.8  &  5$^{+3}_{-2}$  &
70$^{+10}_{-20}$ & 90$_{-60}^{+30}$  &   $-(1.5\div 3.0)\times 10^4$   &  $1512^{+8}_{-5}$ &   0.96$\pm$0.07 (R) & 98/93  \\ 
  1.4$^{+0.5}_{-0.02}$ (V) &  &  &  &  &  &  &   1.2$\pm$0.2 (V) &  \\ 
\hline
\multicolumn{2}{l}{C image}  \\
  2.1$^{+0.7}_{-0.2}$ &  0.97$\pm0.02$  & $10_{-3}^{+10}$   &
85$\pm$5  &  53$\pm$10  &  $(2.0\pm 0.2)\times 10^4$  & 1309$\pm$4  &  2.09$\pm$0.05  &  88/75  \\ 

\end{tabular}
\end{table*}

An alternative to a single fold caustic amplifying a relativistic
strongly-inclined disc is a more complex structure of several caustics
or cusps. We find excellent agreement fitting the amplification curve
with a simplified
disc traversing two fold caustics sequentially in one direction at one
velocity (figure
\ref{fig:twofold}). Best-fit parameters are, for fixed $a=0.2$ and
$M=8.8\times 10^8\Msun$:
$v_{eff}=(1.35\pm0.22)\times 10^4 \kms$, $t_{01} = 2451440\pm 12 \rm d$ for
the first fold and $t_{02} = 2451810\pm 30 \rm d$ for the second, the
ratio of caustic
strengths $l=0.84\pm 0.06$, $\chi^2 = 160/244(118)$. The strengths of
the two caustics are about 0.7 and 0.6, the weaker one is delayed by
$\Delta t = 367\pm 8$~days. The worst problem
for this interpretation of the observational data is the high required
transverse velocity. Spatial  size of the disc is $\propto M^{2/3}$,
and the mass should be sufficiently decreased to match the required
transverse velocity to the observed peculiar velocity distribution for
galaxies. Setting $M=3\times
10^8\Msun$ allows to obtain a reasonable fit ($\chi^2 \simeq 162$) for a
relatively small transverse velocity $v_{eff}\sim 2\times 10^3\kms$. 

\subsection{\einc}\label{sec:res:einc}

Simplified disc fits appear even worse for the case of
\einc. Qualitatively, the situation is understandable: plain disc
symmetric brightness distribution is unable to reproduce the
two-peak structure of image A event as well as the narrow peak
feature of the image C event. Both are relatively well reproduced if
the inclination is high ($i\gtrsim 70^\circ$). 

\begin{figure*}
 \centering
\includegraphics[width=\textwidth]{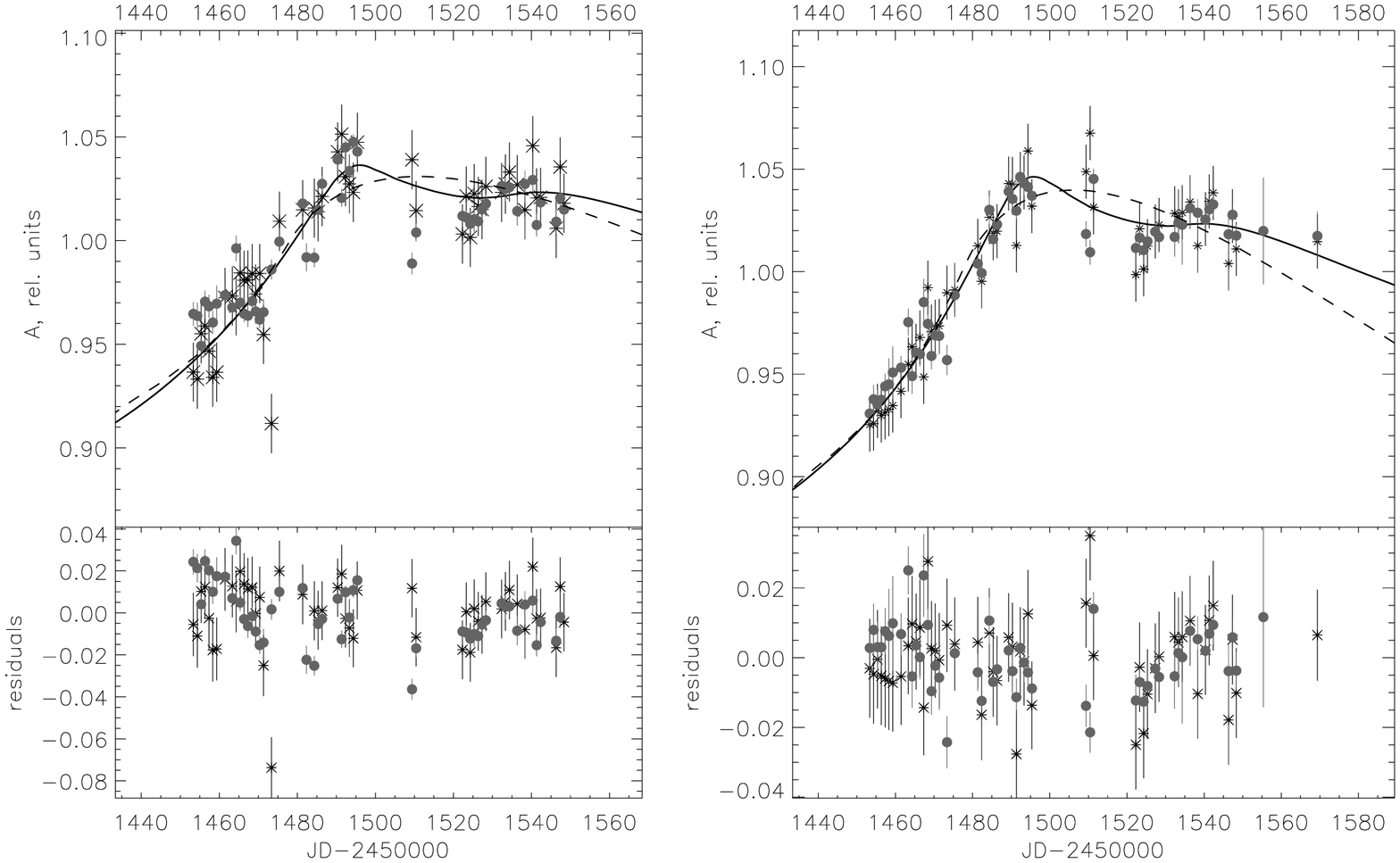}
\caption{Amplification curve for the A image of \einc\ fitted with a
  Kerr disc fold traversal model  with $a=0.2$, $M=7\times 10^9\Msun$,
  $i=70^\circ$, $\psi=96^\circ$. R band is to the left, V band to
  the right. Simplified disc best fit is shown for
  comparison by a  dashed curve. Asterisks and circles correspond to ISIS and
  PSF-fitting fluxes, respectively. 
}
\label{fig:eincA}
\end{figure*}

Running more than 2000 models, we find a reasonable fit for the A image
amplification curve (figure \ref{fig:eincA}). For fitting, we use GLITP data
reduced by the two standard techniques separately. PSF-fitting data have
larger (but probably more reliable) statistical errors hence the minimal
$\chi^2$ is considerably smaller. Higher rotation
parameters (up to $\sim 0.9$) and higher masses (up to $\sim
10^{10}\Msun$) produce apparently worse fits but are
still allowed if one increases the observational uncertainties by a
factor of several. Inclination $i$ and positional angle $\psi$ are
well constrained, because the notable two-peak structure requires
certain disc orientation (see figure \ref{fig:animkerr}). 
At the moment we are satisfied with the qualitative
agreement because the data definitely have some additional
error sources possibly connected to the bad visibility of the object near the
amplification curve maximum. 
Amplification curve shape is similar for both
filters, even the two maxima seen in the R-band curve are also
visible in the V-band. 

\begin{figure*}
 \centering
\includegraphics[width=\textwidth]{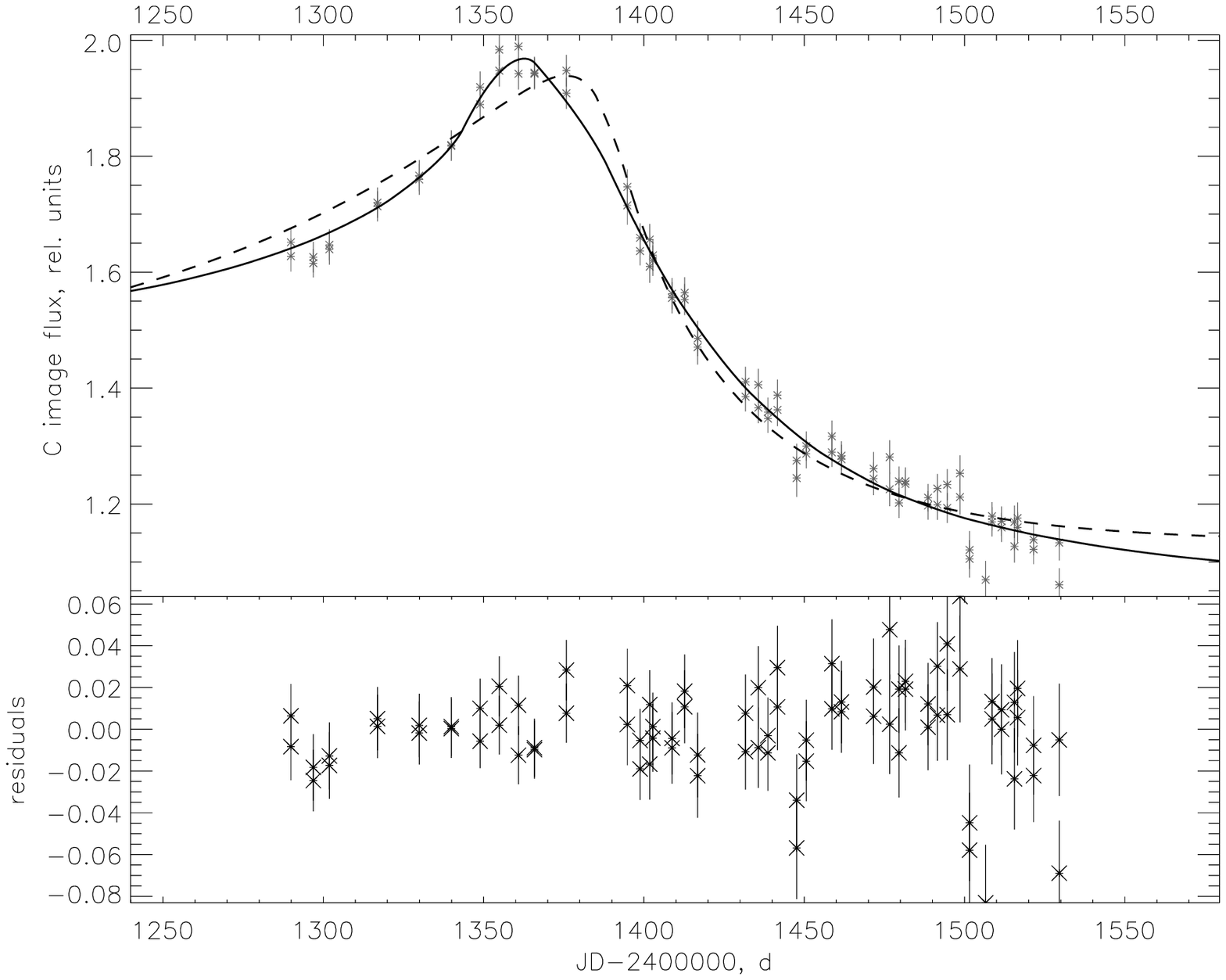}
\caption{Best fit for the \einc\ image C high amplification event, V
  band. Simplified disc fit is shown by a dashed line, solid line is
  the best-fit Kerr metric disc fit ($a=0.97$,  $M=10^{10}\Msun$, $i=85^\circ$
  and $\psi=53^\circ$).
}
\label{fig:eincC}
\end{figure*}

It is impossible to explain the observed amplification curve of the C
image without a relatively large $a\gtrsim 0.95$. The rapid rise of the
curve in figure \ref{fig:eincC} can not be reproduced unless the
accretion disc has a bright spot naturally explained by high rotation
parameter, high mass models. The predicted mass is very high, $M
\gtrsim 7\times 10^9\Msun$. If the C-image amplification is a {\it bona
fide} caustic crossing event, then virial mass measurements
underestimate the masses of quasars by a factor of several. 

\section{Discussion}\label{sec:disc}

\subsection{What do caustic crossings really probe for?}

The fine structure of high-amplification events in the light curves of
microlensed quasars is an effect naturally expected in the framework of
thin accretion disc model. Replacing accretion disc brightness distribution
by some symmetrical structure with brightness uniformly decreasing with
distance would produce a light curve with a single nearly symmetrical peak
with the  duration time scale $r_d/v_{eff}$ defined by the size of the emitting
region (see for example the analytical solutions given in \citet{SW87},
Appendix B). Its shape near the maximum is roughly parabolic. In this case,
the amplification curve maximum is always convex unless the brightness
distribution has a non-monotonic or asymmetric shape. 
Interpreting individual humps in the amplification curves as
individual caustic crossing events leads to unphysically
large transverse velocities several times larger than the upper limits
estimated in section \ref{sec:base}. 

As it was shown in section \ref{sec:plaindisc}, the only parameter
affecting the amplification curve shape in the simplified disc case
is $X = r_d/r_{in}$. Disc size scale itself is degenerate with the effective
transverse velocity. Amplification
curves are primarily sensitive to the combinations $r_{in} /
v_{eff} \sqrt{J}$ and $r_d / v_{eff} \sqrt{J}$ (see section
\ref{sec:plaindisc}) having the physical
meaning of traversal times for the central hole in the disc and the
disc radial scale itself, respectively. This
also holds for the general relativistic case as long as the disc is
nearly face-on.

Relativistic effects break these degeneracies at
inclinations $i\gtrsim 60^\circ$. 
Amplification curve properties become sensitive to the
two angles $i$ and  $\psi$, but the dependence on $a$ is weaker and
similar light curves emerge for similar values of $X$ if inclination
and positional angle are the same. 


Interpretation of the three events relies on their consideration as
caustic-crossing events. For \einc, this is justified by the works of
\citet{GGG06,koptelova07} who performed detailed simulations to check whether
artificial amplification maps produce caustic crossings similar (in
some sense) to the observed peaks in the light curves of individual
images. 

For the case of \sbs, optical depths to microlensing are relatively
small. Central convergence of $\kappa_c = 0.5$ sets an absolute upper
limit for it. The optical depth to microlensing is smaller than
$\kappa_c$ unless the stars in the optical path are strongly clumped,
that is unlikely
for an elliptical galaxy (probability of accidentally hitting a globular
cluster is $\sim 10^{-5}$). If dark matter makes significant
contribution to the gravitational field of the lensing galaxy, optical
depths should be decreased proportionally toward the probable value
of $\sim 0.1$ for the stronger lensed B image. Image A is about three
times farther from the centre of the lens that makes it much worse
candidate for microlensing. 
The caustic net structure in this case is determined
by small shears distorting single-lens amplification
patterns \citep{kofman}. 
Therefore, there is a strong probability for passing a close
caustic pair. However, in this case the caustics are traversed in
opposite directions that should produce a pattern completely different
from the observed amplification curve. 


In this work, we used Kerr-metric geodesics and relativistic Doppler
effects, but the disc model was not fully
relativistic. We used the temperature law of \citet{SS73} instead of
the more accurate 
relativistic thin disc law provided by \citet{NT73,PT74}.
Difference in the two laws is in the correction factor that
decays more smoothly inwards and is generally smaller for the fully relativistic
thin disc case. We tested the difference between the amplification curves
calculated using these two temperature laws and found that for
relevant black hole parameters ($M\lesssim 10^{10}\Msun$ and $a>
-0.9$), the difference is smaller than one percent of the peak
amplification. It is more important to consider the
effects of non-trivial inner boundary condition, shock formation and
rapid disc tilt change in the inner parts of the disc and then pass to finer
effects like the general relativity corrections to the temperature law. 

\subsection{Caustic strengths}

For \sbs, amplification curve amplitude is moderate and may be
characterised by a caustic strength of $k = \mu_1 / \mu_0 \sim
0.6$. Using the estimates made by \citet{WKR93}, one may expect a
mean caustic strength of:

\begin{equation}\label{E:kstrength}
\langle k \rangle \sim 0.56 \left(\langle m
\rangle / \tau \right)^{1/4},
\end{equation} 

where $\tau \simeq 0.1\div 0.5$ is
the optical depth to microlensing, and $\langle m\rangle \sim 0.1\div
0.5$ is the
mean microlens mass in Solar units. Note that the dependence is
very weak both on the optical depth and on the mean mass thus making
the caustic strength a good consistency check for the model.
In the case of \sbs, observational data
argue for $\langle m \rangle / \tau \sim 1$, consistent with a
moderately sub-solar mean stellar mass. Low end of the mass
function of red and brown dwarfs is poorly known and probably variable
\citep{bastian_review}. There are indications for a low mean
stellar/substellar mass ($\langle m \rangle \sim 0.1$) in the old
stellar population of elliptical galaxies
\citep{dokkum_natura}, therefore the optical depth to microlensing is
most likely relatively small, $\tau \sim 0.1$.
If the amplification curve for \sbs\ is interpreted as a superposition
of two caustic crossings, the two folds have similar strengths of 0.7 and
0.6, consistent as well with a stellar population rich in low-mass
objects.

For \einc\ events, the amplitudes are higher. Qualitatively, it was
expected, because the lens is a spiral galaxy and the number of
massive stars among the microlenses should be higher, with the mean
mass of $\sim 1\Msun$. Mean caustic strength $\langle k
\rangle\simeq 2$ (as for the image C event) only if $\langle m\rangle /\tau \sim 
100\Msun$ and hence $\langle m\rangle \sim 40\Msun$ (numerical simulations suggest that the optical depth to microlensing is
$\tau\sim 0.4$  for Einstein's cross \citep{bate11}) that is considerably
higher than what one
may expect even from a young stellar population. 
In most simulations with $\tau \sim 0.5$ (see for example
\citet{SW02,SW04}), predicted amplification factor
distributions are broad and allow
caustics with strengths exceeding the mean value of $k$ by a factor of $\sim
2$. The reason why our estimates do not contradict the
theory is selection effect: we consider two brightest and most evident
high amplification events while the total number of fainter events
present in the light curves of the four images in the about three-year
interval of OGLE-II monitoring program is about 10 (if the mean
distance between fold traversals is about a year). 


\subsection{Black hole masses and disc sizes}\label{sec:disc:mass}

The relative size of the inner disc hole and the observed strength of
relativistic effects favour black hole masses significantly higher
than virial estimates. For \sbs\ and \einc, the virial masses
estimated using the width of the CIV$\lambda$1549
resonance line are equal to $8.8\times 10^8$ \citep{peng06} and
$9\times 10^8\Msun$ \citep{YdR},
correspondingly, with the proposed uncertainty of about a factor of
2. Single caustic crossing event fitting argues for considerably higher
masses, $(2\div 3)\times 10^9\Msun$ and $(7\div 9)\times 10^9\Msun$. 

Altering black hole masses by a factor of $\sim 3$ was proposed by
\citet{morgan10} to explain the inconsistency between the accretion
disc sizes estimated using microlensing statistics and the accretion
disc size predictions following from the standard disc theory. Note
that in our study the degeneracies are different from those in \citet{morgan10}.
While the Monte Carlo light curve analysis is sensitive to the disc size
itself ($r_d \propto \left(l/\eta\right)^{1/3} M^{2/3}$), the shapes
of caustic-traversal events primarily reflect the ratio of the disc radial
size scale to its inner boundary size ($X\propto
\left(l/\eta\right)^{1/3} M^{-1/3}$). To reproduce our results without
altering black hole masses we should decrease the Eddington ratio by a
factor of several rather than accretion efficiency as it was proposed by
\citet{morgan10}. 


To distinguish between higher black hole masses and lower $l/\eta$ ratios, one
may use either photometrical data or different-technique accretion disc size
estimates. 
Central black hole mass may be estimated using amplification-corrected magnitudes. 
Observed flux from a thin disc is calculated by integrating the observed
intensity (indices ``obs'' and ``em'' refer here to the radiation intensity in
the observer frame and in the frame comoving with the QSO):

$$
\begin{array}{l}
F_\nu = \int I_{\nu,obs} d\Omega = \\
\qquad{} = \frac{2\pi}{D^2\times(1+z)^3} \cos i \times  \int I_{\nu,em}\left(\nu/(1+z)\right) R dR,
\end{array}
$$

where $D=D(z)$ is angular size distance. This distance scale is known for its
non-monotonic dependence on redshift. Maximal distance of $ 1.7\Gpc$ is
reached at
the intermediate redshift of $z\simeq 1.6$ for standard $\Lambda$CDM. Most of
the well-studied lensed quasars have redshifts close to this value.
 
For $I_\nu$, we substitute the Planck law
with the temperature law determined by the standard accretion disc theory as:

$$
T=\left(\frac{3}{2}\frac{G^2M^2}{\sigma \kappa c}\frac{l}{\eta}\right)^{1/4} R^{-3/4}
$$

Here we neglect the correction term that has only small influence on the
integral flux since the area of the disc affected by the term is about
$(r_{in}/2.44 r_d)^2 \lesssim 0.1$ times smaller. For the integration limits set
to 0 and $+\infty$, the integral is reduced to the following \citep{AS72}:

$$
\int_0^{+\infty} \frac{x^{5/3}dx}{e^x-1} = \Gamma(8/3)\zeta(8/3)
$$

Observed flux is finally expressed as:

$$
\begin{array}{l}
F_\nu = \frac{4\pi h\nu_{em}^3}{c^2} \frac{\cos i}{D^2\times (1+z)^3} \int
\frac{RdR}{\exp\left(\frac{h\nu}{kT(R)}\right)-1} = \\
\qquad{} = 8\pi \left(\frac{2}{3}\right)^{1/3} \Gamma(8/3)\zeta(8/3)
\frac{k^{8/3}\nu_{obs}^{1/3}}{c^{8/3}h^{5/3}\kappa^{2/3}\sigma^{2/3}}\times \\
\qquad{} \times
(GM)^{4/3} \left(\frac{l}{\eta}\right)^{2/3} \cos i \times \frac{1}{D^2\times
  (1+z)^{8/3}} \simeq \\
\qquad \simeq 6.9 \cos i \left(\frac{l}{\eta}\right)^{2/3} \left(\frac{\lambda_{obs}}{1\mu}\right)^{-1/3}
\left(\frac{M}{10^9\Msun}\right)^{4/3} \times \\
\qquad{} \times  \left(\frac{D}{1\Gpc}\right)^{-2}
(1+z)^{-8/3} \mJy
\end{array}
$$

This flux may be used to independently estimate the black hole mass.
In particular, for the {\it HST} F814W filter
 (we used the calibration given in \citet{holtzman}; this
allows to use the magnitudes given by \citet{morgan10} in table 1) the mass may be
estimated photometrically as follows:

\begin{equation}\label{E:masses}
\begin{array}{l}
M \simeq 2.7\times 10^7 \left( \frac{D}{1\Gpc} \right)^{3/2} (1+z)^2 \times \\
\qquad{} \times \sqrt{\frac{\eta}{l}} \cos^{-3/4} i \times 
10^{-0.3(I-19)} \Msun\\
\end{array}
\end{equation}

\begin{figure}
 \centering
\includegraphics[width=\columnwidth]{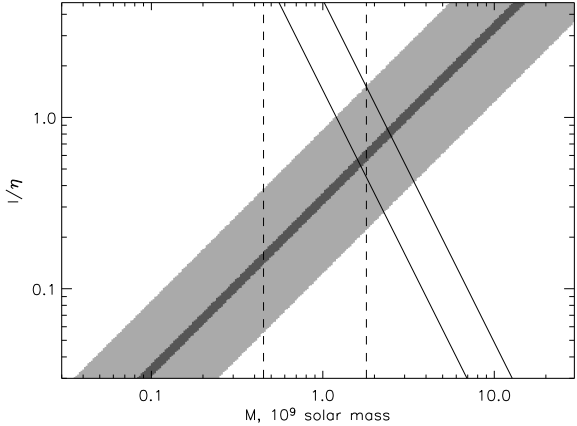}
\caption{Mass and $l/\eta$ for \einc. Shaded region is the band restricted by
  the light curve fitting made in the present work (the darkest region
  corresponds to the intersection of the both mass estimates). Vertical dashed
  lines show the range set by the virial mass estimate with the relative
  uncertainty of $ 0.3\rm dex$. Solid lines show the constraints set by
  equation~(\ref{E:masses}). Disc inclination is set to $60^\circ$. 
}
\label{fig:meleta}
\end{figure}

Following the original work where a similar formula was used to estimate the
photometrical radius, we denote the magnitude by $I$ and use the
amplification-corrected values of $I(\einc)=17.9\pm 0.44\magdot{\,}$ and
$I(\sbs)=18.92\pm 0.13\magdot{\,}$.
Photometrical masses are closer to the virial mass estimates. These estimates
are degenerate with the $l/\eta$ ratio in a way different from
the results of our light-curve analysis that
allows to reach consistency if the Eddington ratio is low, $l/\eta \sim
0.1$. 
In figure \ref{fig:meleta}, we show the estimates for
$l/\eta$ and black hole mass made with different methods. Monte-Carlo
amplification curve fitting is sensitive to $r_d \propto M^{2/3}\times
(l/\eta)^{1/3}$, while the observed amplification-corrected flux scales as $F_\nu
\propto M^{4/3}\times (l/\eta)^{2/3}$ that results in parallel bands in the
graph. For \einc, the microlensing and photometrical radii are consistent
within the uncertainties. Our data may be made consistent with
photometry and virial masses for narrow ranges of parameter values. 
In particular, for \einc, there is a zone where all the three
bands intersect, $M\simeq(1.6\div 1.8)\times 10^9 \Msun$ and $l/\eta
\simeq 0.5\div 0.6$.
 In the case of \sbs, photometry, virial and light-curve fitting
intersect for $M\simeq(0.8\div 1.1)\times 10^9\Msun$ and $l/\eta \simeq
0.7\div 1.3$. However, in this case the microlensing radius found by
\citet{morgan10} is considerably higher.

Consistency between different methods of radius and mass estimates allows to
suggest that the microlensing radii found by \citet{morgan10} are a subject to
some bias that
increases in some cases the effective radius by a factor of several. This
may be
contamination from a source of much larger angular size (such as unresolved
starlight or the broad-line region) or some optically-thin scatterer. These
effects were considered in the original work and seem to qualitatively explain
the observed discrepancy. 

Since in general there are inconsistencies between the accretion disc radii
obtained by different methods, we refrain from any final conclusions upon the
masses of the
central black holes in QSO discs. Still it seems plausible that the apparently
large disc sizes may be connected to larger black hole masses, smaller
Eddington ratios and high inclinations (the issue of these will be considered
in detail in the next subsection). We also do not exclude that some
non-standard accretion regimes may influence the observational properties of
quasar accretion discs. In particular, low accretion rates may lead to an
optically-thin advection dominated region in the inner parts of the disc. The
innermost stable orbit radius will be over-estimated in this case resulting in
black hole masses and accretion disc sizes biased toward higher
values. We will consider the effects of non-standard accretion regimes
on microlensing variability in subsequent papers.

Larger black hole masses mean also larger disc sizes. For black hole masses
$M\sim 10^{10}\Msun$, disc size in the considered wavelength range becomes
$\sim 10^{16}\rm cm$ that is comparable to the Einstein radius $r_{Ein}$ in
the case of \sbs. Straight caustic approximation is violated in this case
because $\theta_{Ein}$ also sets the mean curvature radius of fold
caustics as well as the mean distance between individual folds for
intermediate $\tau$ \citep{GP02}.
For the case of \einc, Einstein-Chwolson radius
is larger because of the smaller distance $D_L$. Most of the results of this
work refer to the inner parts of the disc that makes the effects of caustic
curvature even smaller. For future more accurate studies, we propose using
the more precise parabolic model \citep{FW99}.

\subsection{Inclinations of QSO discs}\label{sec:disc:tilt}

High inclinations ($i\gtrsim 70^\circ$) are required if we
interpret the three considered high amplification events as caustic
crossings. All the three events show fine structure near their
maxima. It is possible to explain this by highly inclined,
relativistic discs around black holes with masses considerably higher
than the virial estimates. One amplification curve (\einc, image C)
allows to restrict both the Kerr parameter and the mass to a high
accuracy. Still it should be noted that these estimates are sensitive
to the inner structure of the disc and therefore model-dependent. 
High inclinations are needed because the observational data require
asymmetric brightness distributions.  

Still, there are strong reasons why QSO discs {\it should not} be inclined by
$\gtrsim 70^\circ$:

\begin{itemize}

\item Outer disc rim has a non-negligible thickness and is able to
  obscure the central UV-radiating parts if inclination is $i \gtrsim
  \pi/2 - h/R$, where $h/R \sim 0.02$ is the relative disc thickness
  \citep{SS73}; this is important if the disc has very high
  inclination $i \gtrsim 85^\circ$.

\item A stricter limit is set by dust tori detected
  spectrally for most quasars. Hot dust emission is observed from
  \einc\ \citep{agoldust} consistent with a large-scale ($R \gtrsim
  0.03\pc$) structure of dust intercepting considerable part of the
  quasar luminosity and re-radiating the absorbed luminosity in the
  infrared range.

\item A disc observed edge-on should appear under-luminous. For a thin
  disc, the observed luminosity is $\cos i$ times lower than for the
  face-on case, for a concave disc the decrease is stronger. Thus the
  real luminosity should be a factor of $\cos^{-1}i \sim 10$
  higher. In the case of \einc, a strong observational evidence for small
  inclination ($i>60^\circ$ is ruled out at a 95\% confidence level)
  is given in \citet{eincincl}.

\end{itemize}

\begin{figure}
 \centering
\includegraphics[width=\columnwidth]{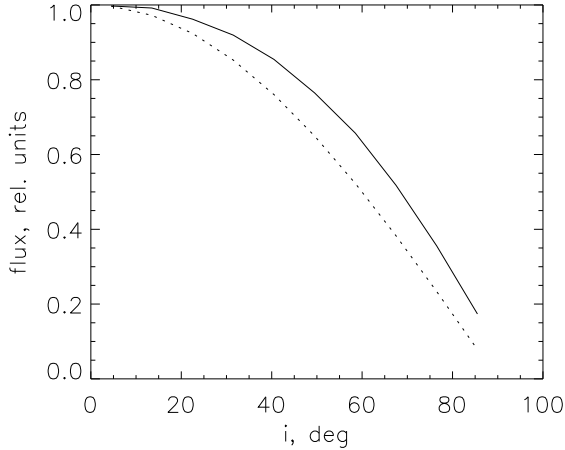}
\caption{Flux dependence on inclination for an accretion disc around
  an $M=10^9\Msun$, $a=0.6$ black hole. Dotted line shows the $\cos i$
  law for comparison. 
}
\label{fig:oridisc}
\end{figure}

The strength of the last argument is diminished by the high inferred
masses of the central black holes. An order of magnitude higher black
hole mass leads to an order of magnitude higher Eddington
luminosity. This makes the observed flux consistent with the
microlensing disc size estimates. Besides, \citet{eincincl} did not
consider relativistic effects that have the potential to distort the
dependence of the observed flux on inclination significantly. For the
moderate black hole mass of $10^9\Msun$ and $a=0.6$, luminosity
dependence on inclination deviates by more than 20\% at $i>40\deg$ and
by more than 50\% at $i>70\deg$ (see figure \ref{fig:oridisc}).

An interesting clue is that \sbs\ is a BAL (broad absorption line)
quasar \citep{chavushyan97}. Though BAL QSO are generally considered
as seen nearly edge-on \citep{elvis00}, some of these objects have
strongly variable bright radio emission suggesting a collimated
relativistic outflow observed at low inclination \citep{zhou06}. It is
still an open question how one can collimate a jet in a direction
different from the normal to the disc plane. 

A possible solution is that the angular momenta of the
accreting black holes in \sbs\ and \einc\ are misaligned with the angular
momentum of the infalling matter. 
Since there are no reasons for a supermassive black hole to be
aligned with the accreting matter at large distances, this
is probably the case. The inner parts of the flow should be tilted and
warped. Most investigations predict alignment of the inner parts of
the flow known as Bardeen-Petterson effect, see \citet{bardeenpetterson},
normally inside the $R\sim 10\, r_{in}$ region where the observed UV
radiation is emitted. Similar misalignment may be proposed a
possible reason for the apparent inconsistency between the
inclinations inferred from radio and optical properties of radio-loud
BAL quasars. 

\begin{figure}
 \centering
\includegraphics[width=\columnwidth]{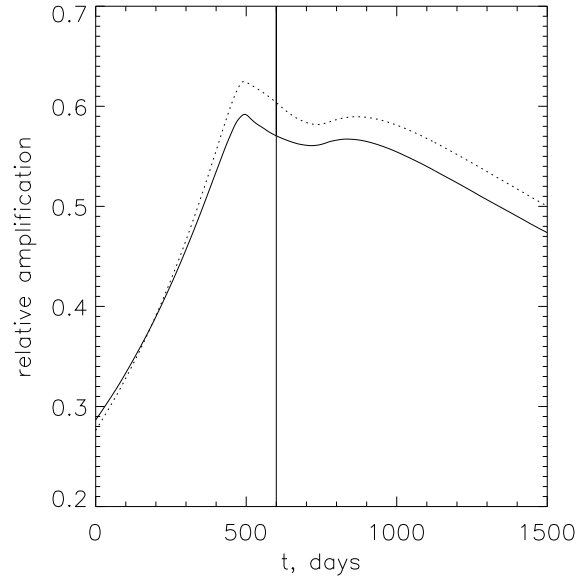}
\caption{
Simulated amplification curves of an inclined (solid line) and tilted
(dotted) discs. For both curves, $a=0.9$, $M=10^9\Msun$, maximal inclination is
80$^\circ$, $\psi\simeq 0$, $v_{eff}=-10000\kms$. See section
\ref{sec:disc:tilt} for details. 
}
\label{fig:inclineddiscs}
\end{figure}

In figure \ref{fig:inclineddiscs} we show the  predicted amplification
curve for a thin accretion disc with zero
inclination at infinite distance but tilted toward the plane of
the black hole inclined by $\theta_{bh}$ to the picture plane (and to the
plane of the disc at large distances). For inclination, we use the law
$\theta(r) = \theta_{bh} \cdot \exp\left(-r/r_{in}\right)$. Rotation
law for an inclined disc was taken from \citet{shakura87}, while the
temperature law was assumed standard. Every annulus was considered
flat that is of course an over-simplification but still serves the
main goal to show that fold caustic traversal curves are primarily
sensitive to the innermost parts of the accretion disc. 

\subsection{Influence of disc tilts upon the unification scheme}

In case of random orientation of the black hole spin and the orbital
angular momentum of the matter entering the disc, an average disc will
have a tilt of $\theta \sim 90^\circ$ at large distances. Half of
the discs should be counter-rotating. In case of random
misalignment, half of the disc planes should have inclinations larger
than 60$^\circ$ with respect to the black hole rotation frame.  
A tilted disc around a Kerr black hole is usually proposed to
gradually change its inclination with decreasing radius approaching
alignment at the innermost stable orbit
  Linear analysis made by  \citet{slava} for a
thin disc around a nearly-Schwarzschild black hole confirms this
conclusion only for the case of considerably large viscosity (see the
original work for
details). The authors also show stability of counter-rotating discs
around slowly rotating black holes. 
On the other hand, numerical simulations made by
\citet{fragile_anninos,dexter_fragile} suggest that the disc evolution
toward alignment should take place in its innermost parts, at several
Schwarzschild radii.

Tilted discs may become a serious obstacle for formation and
observability of relativistic jets. Once formed, relativistic outflows
will be stopped by the higher-pressure material of the disc
or the dust torus. 
If jet formation and propagation into the rarefied galactic medium becomes
impossible in case of strongly inclined discs, one may try
to explain the observed radio brightness dichotomy of QSO by
different regimes of alignment, as it was proposed by \citet{LE10} for
Seyferts and radiogalaxies. Presence of a thick torus with
$\theta_t \sim 30^\circ$ leads to about 75\% probability that the
jet will be stopped either by the torus or by the tilted disc itself. The real
fraction of radio-weak sources is higher, about 90\% (see for example \citet{LBQS7}).

Importance of disc tilts for jet formation may also have connection to
the formation of the ``localised'' X-ray radiation found in some
QSO. In particular, in the two recent works by \citet{zimmer} and
\citet{einc_xrays}, the properties of the X-ray
emitting region in \einc\ are recovered using microlensing curves. In
\citet{einc_xrays}, it is concluded that this region is smaller and more
variable than any existing extended jet or corona model can explain.
Among other possibilities, the ``failed jet'' model by
\citet{ghisellini} is put forward. If the jet possesses
enough energy to escape the gravitational
well but is stopped by the disc matter, the source of the X-ray
emission may be connected to the hot spot in the disc where its energy
is dissipated. Relativistic termination
shock remains invisible but the bow shock where the ram pressure is
balanced by the pressure of the disc should make important contribution
to the observed X-ray emission. And the spatial size of the X-ray
bright spot may be small if the jet is sufficiently collimated.
 Stopped jet
model may be checked by considering correlations between different
properties of X-ray and radio emission from quasars. 

\section{Conclusions}

We show that general relativity effects have considerable influence on
the amplification curves of microlensed quasars. This effect is more
profound if the disc is strongly inclined. Having better data on hand
(smaller observational errors, larger homogeneity and better temporal coverage)
may allow to
resolve the structure of high-amplification events with a better precision
and to use them for accurately measuring disc tilts and black hole
rotation parameters of lensed quasars. 

Some implications may be made even now. In particular, the discs in
both objects, \sbs\ and \einc, are seen at high inclinations. 
Apparent contradiction with the high observed fluxes and lack
of strong absorption signatures in the spectra may be resolved if one
considers a disc having low inclination ($i\lesssim 60^\circ$) at
tens of $GM/c^2$ but is seen nearly edge-on in its inner
parts. Similar picture is expected if the initial angular momentum of
the accreted matter is strongly misaligned with that of the black
hole. 

High probability to effectively measure a high inclination may be
qualitatively
understood as a very strong tilt and warp near the last stable
orbit. In this case, there should be, at a high probability, some
radius at which the disc lies perpendicular to the picture
frame. Contribution of this annulus should be enhanced by Doppler
boost. In other words, if a disc has a range of inclinations
due to tilts and warps, higher inclinations will make larger
contribution to microlensing amplification curves.

In the analysed data we do not find any challenges for the
standard accretion disc model. But they bring our attention toward the
role of disc tilts that was never considered in the framework of QSO
microlensing. Besides, some of accretion disc properties are better
  understood if we alter the accretion disc parameters such as Eddington ratio
  and accretion efficiency. The apparently high black hole masses and disc
  sizes required by microlensing effects may be connected to the low accretion
  efficiency and contamination by larger angular scales.

\section*{Note added in proof:}

We became aware of the papers by \citet{BC02,BC04} where amplification curves
considered in this work were analysed. Authors restore one-dimensional
brightness distributions and propose to connect their shapes to relativistic
effects. 

\section*{Acknowledgements}

The article makes use of observations made with the Nordic Optical
Telescope, operated on the island of la Palma jointly by Denmark,
Finland, Iceland, Norway, and Sweden, in the Spanish Observatorio del
Roque de los Muchachos of the Instituto de Astrof\'{\i}sica de
Canarias. It also uses the public data from OGLE-II and GLITP
archives. We would like to thank Ilfan Bikmaev and Irek Khamitov for
the RTT data on \sbs, and Elena Shimanovskaya and Boris Artamonov for their
help with the
photometric data on \einc. We acknowledge the use of RFBF grant
09-02-00032-a and thank Max Planck Institute for Astrophysics (MPA
Garching) for its hospitality. Special thanks to R. A. Sunyaev
for valuable discussions. We are also grateful to the referee who helped us
to improve the quality of the work. 

\bibliographystyle{mn2e}
\bibliography{mybib}

\label{lastpage}

\end{document}